# Engineering topological states and quantum-inspired information processing using classical circuits

Tian Chen, Weixuan Zhang, Deyuan Zou, Yifan Sun and Xiangdong Zhang[*]

*Key Laboratory of advanced optoelectronic quantum architecture and measurements of Ministry of Education, Beijing Key Laboratory of Nanophotonics & Ultrafine Optoelectronic Systems, School of Physics, Beijing Institute of Technology, 100081, Beijing, China*

[*]*Author to whom any correspondence should be addressed. E-mail: zhangxd@bit.edu.cn*

**Abstract:** Based on the correspondence between circuit Laplacian and Schrodinger equation, recent investigations have shown that classical electric circuits can be used to simulate various topological physics and the Schrödinger's equation. Furthermore, a series of quantum-inspired information processing have been implemented by using classical electric circuit networks. In this review, we begin by analyzing the similarity between circuit Laplacian and lattice Hamiltonian, introducing topological physics based on classical circuits. Subsequently, we provide reviews of the research progress in quantum-inspired information processing based on the electric circuit, including discussions of topological quantum computing with classical circuits, quantum walk based on classical circuits, quantum combinational logics based on classical circuits, electric-circuit realization of fast quantum search, implementing unitary transforms and so on.

# Contents

# 1. Introduction

In the modern information era, how to improve the ability of signal processing is a common interest in the fields of science and engineering. Quantum information process is believed to solve certain problems substantially faster than traditional computing methods, which has attracted much attention in recent decades [1-5]. However, the quantum information process schemes also face two bottlenecks: scalability and decoherence [1]. Although there have been significant progresses in the research of quantum information process in recent years [6-14], the wide applications within these quantum schemes still keep great challenges.

On the other hand, recent investigations have shown that classical electric circuits can be used to simulate various topological physics [15-17] and the Schrödinger's equation [18-20] due to the correspondence between circuit Laplacian and lattice Hamiltonian. Motivated by these studies, a series of research on quantum-inspired information processing based on electric circuits have been implemented. In this article, we review the recent developments of topological physics based on classical circuits and quantum-inspired information processing using classical circuits.

In Section 2, the relation between the circuit Laplacian in the electric circuit and the Schrödinger equation in the quantum system is shown explicitly. Then, in Section 3, a short review about the topological phenomena based on the classical circuits is presented. The topological physics in Hermitian and non-Hermitian systems are discussed, and extended to the observation of the topological order. Since the realization of classical analog of topological order has an important application in the topological quantum computation, we review the works about the topological quantum computation with classical circuits in Section 4. Not only in topology, the electric circuit can be employed to realize the quantum walk and its related quantum algorithms. The corresponding quantum algorithms on the circuit network are shown in Section 5. In addition, based on the quantum walk in the circuit, the quantum combinational logics and their applications in quantum speedup algorithms have been reviewed in Section 6. In Section 7, quantum-inspired microwave signal processing for implementing unitary transforms has been reviewed. The conclusion and outlook are given in Section 8.

## 2. The general theory on the correspondence between circuit networks and quantum lattice models

In this section, we elaborate on the general theory for the correspondence between circuit networks and quantum lattice models. We begin by introducing the mathematical correspondence between the circuit Laplacian and the Schrödinger equation. Following this, we detail the methods for implementing various couplings in circuit networks, including the complex couplings for simulating magnetic fields, non-reciprocal couplings for the realization of non-Hermitian effects, and the incorporation of nonlinear on-site potentials, as well as time-varying couplings.

**2.1 Analog of circuit Laplacian and Schrödinger equation**

An electrical circuit network can be represented as a graph, with nodes and edges corresponding to circuit junctions and connecting wires or elements, respectively. Consider a circuit network composed of resistances, capacitances, and inductances, according to Kirchhoff's law, the $N$-component vectors $\vec{I}$ and $\vec{V}$ which indicate the current and voltage at the nodes of the circuit can be expressed as:

$$\frac{d}{dt}\vec{I}(t) = \frac{d^2}{dt^2}\Gamma\vec{V}(t) + \frac{d}{dt}\Sigma\vec{V}(t) + \Lambda\vec{V}(t), \tag{1}$$

where the $\Gamma$, $\Sigma$, and $\Lambda$ are the $N \times N$ matrices, and the elements in these matrices represent the capacitances, the inverse of resistances, and the inverse of inductances between two nodes or node to ground [21].

The analogy of the time-dependent Schrodinger equation $i\partial_t |\psi(t)\rangle = H |\psi(t)\rangle$ is initially introduced [22, 23]. The Resistance-Capacitance (RC) circuit is widely used to analog the time-dependent Schrodinger equation. Considering a RC circuit where capacitance only exist in the ground parts, according to Equation (1), the current and voltage at the nodes $j$ of the circuit can be expressed as:

$$i\frac{dV_m}{dt} = \frac{i}{C_j}(-\frac{1}{R_{m0}} - \sum_{\langle n \rangle}\frac{1}{R_{mn}})V_j + \frac{i}{C_j}\sum_{\langle n \rangle}(\frac{1}{R_{mn}}V_n), \tag{2}$$

where $V_m$ and $V_n$ are the voltages of the node $m$ and $n$. $C_{m0}$ and $R_{m0}$ represent the capacitance and the resistance between node $m$ and ground. $R_{mn}$ represent the resistance between node $m$ and $n$. $\langle n \rangle$ indicates the summation confined to other connected nodes. It can be reformulated in matrix form

$$i\partial_t |\phi(t)\rangle = H_T |\phi(t)\rangle, \tag{3}$$

where $|\phi(t)\rangle = (V_1(t), \cdots, V_N(t))^T$ and

$$H_T = \frac{i}{C_j} \begin{pmatrix} -\frac{1}{R_{10}} - \sum_{\langle n \rangle} \frac{1}{R_{1,n}} & \frac{1}{R_{1,2}} & \cdots & & \frac{1}{R_{1,N}} \\ \frac{1}{R_{2,1}} & \cdots & \cdots & & \cdots \\ \cdots & \cdots & \cdots & & \frac{1}{R_{N-1,N}} \\ \frac{1}{R_{N,1}} & \cdots & \frac{1}{R_{N,N-1}} & -\frac{1}{R_{N0}} - \sum_{\langle n \rangle} \frac{1}{R_{1,n}} \end{pmatrix}. \qquad (4)$$

Comparing Equation (3) with the time-dependent Schrodinger equation $i\partial_t |\psi(t)\rangle = H|\psi(t)\rangle$, the $H_T$ corresponds to the Hamiltonian $H$, and the voltages $|\phi(t)\rangle = (V_1(t), \cdots, V_N(t))^T$ correspond to the eigenstates $|\psi(t)\rangle$ of the lattice model.

We next introduce the analogy of stationary Schrodinger equation $H|\varphi\rangle = E|\varphi\rangle$ [17]. Considering the monochromatic frequency response, the current and voltage in Equation (1) can be express as

$$I(t) = e^{i\omega t}$$

$$V(t) = e^{i\omega t}. \qquad (5)$$

When substituting Equation (5) into Equation (1), the current and voltage at the nodes of the circuit can be expressed as:

$$I_m(\omega) = \left(i\omega C_{mn} + R_{mn}^{-1} + (i\omega L_{mn})^{-1}\right) \sum_{\langle n \rangle} (V_n(\omega) - V_m(\omega)) - \left(i\omega C_{m0} + R_{m0}^{-1} + (i\omega L_{m0})^{-1}\right) V_m(\omega), \qquad (6)$$

where $I_m$ and $V_m$ are the net current and voltage of the node $m$ with the angular frequency $\omega$. $C_{mn}$, $R_{mn}$ and $L_{mn}$ represent the capacitance, resistance, and inductance between nodes $m$ and $n$. $C_{m0}$, $R_{m0}$ and $L_{m0}$ represent the capacitance, resistance, and inductance between node $m$ and ground. $\langle n \rangle$ indicates the summation confined to other connected nodes. It can further be written as

$$I(\omega) = J(\omega)V(\omega), \qquad (7)$$

where $J(\omega)$ represents the admittance matrix of the circuit. By properly choosing the grounding elements (capacitance, resistance, inductance) and assuming that there are no external sources,

Equation (6) can be written in a matrix form as:

$$\beta(\omega)\begin{pmatrix} V_1 \\ ... \\ V_N \end{pmatrix} = \tilde{H}\begin{pmatrix} V_1 \\ ... \\ V_N \end{pmatrix}. \tag{8}$$

Through a reasonable design of the couplings between nodes and node to the ground, $\tilde{H}$ can be made frequency free, and $\beta(\omega)$ is a function of frequency. Comparing this with the Schrödinger equation $H|\varphi\rangle = E|\varphi\rangle$, the $\tilde{H}$ corresponds to the tight-binding Hamiltonian $H$, where the $\beta(\omega)$ and the voltage $V_i (i=1,2,...N)$ correspond to the eigenvalues and eigenstates of the lattice model. The on-site potentials and hopping terms in the lattice model are represented by the diagonal and off-diagonal elements in $\tilde{H}$, respectively. Therefore, properties of quantum systems can be analyzed through voltage and frequency measurements in carefully designed circuit networks.

Besides, if the expression of a designed circuit cannot be written in a matrix form as Equation (8), but can only be written as

$$J(\omega)\begin{pmatrix} V_1 \\ ... \\ V_N \end{pmatrix} = 0, \tag{9}$$

the properties of quantum systems cannot be analyzed through the eigen-frequency. In this case, only at a certain frequency $\omega_0$, the circuit Laplacian $J(\omega_0)$ corresponds to the Hamiltonian $H$. The properties of quantum systems can be analyzed by reconstructing the admittance spectrum using voltage measurement and Fourier transform. The eigenvalues and eigenstates of reconstructed circuit Laplacian $J(\omega_0)$ correspond to the eigenvalues and eigenstates of the lattice model.

## 2.2. Implementation of Various Complex Couplings in Circuit Networks

The above section illustrates the correspondence between circuit networks and quantum lattice models, wherein circuit components such as resistors, inductors and capacitors are employed to simulate fundamental couplings in quantum lattices. However, there are also other unconventional topological phase beyond the fundamental couplings. In this section, we will review the literature on circuit realization of unconventional topological phase, encompassing circuits involving phase factors, non-reciprocal characteristics, negative effects, nonlinear phenomena and time-varying properties.

**2.2.1 Circuits involving phase factors**

In this part, we review the realization of circuits involving phase factors. To simulate hopping terms with phase factors, we often use multiple circuit nodes to represent a lattice site, where these internal degrees of freedom can be used to construct circuit pseudospins with effective coupling phases. The value of coupling phases for the voltage pseudospins depends on the number of circuit nodes acted as a single lattice site. For instance, a hopping term $e^{i2\pi/3}$ can be fulfilled by applying three circuit nodes as a single site, as shown in Figure 1a [24]. Three subnodes are connected by capacitances $C$ and a pair of node groups are crossly connected through three inductances $L$. In addition, each site is also grounded through capacitances $C_g$. The Kirchhoff equation for this setup can be written as:

$$\begin{bmatrix} I_{a,1} \\ I_{a,2} \\ I_{a,3} \end{bmatrix} = i\omega C_g \begin{bmatrix} 1 & 0 & 0 \\ 0 & 1 & 0 \\ 0 & 0 & 1 \end{bmatrix} \begin{bmatrix} V_{a,1} \\ V_{a,2} \\ V_{a,3} \end{bmatrix} + i\omega C \begin{bmatrix} 2 & -1 & -1 \\ -1 & 2 & -1 \\ -1 & -1 & 2 \end{bmatrix} \begin{bmatrix} V_{a,1} \\ V_{a,2} \\ V_{a,3} \end{bmatrix} - \frac{i}{\omega L} \begin{bmatrix} V_{a,1} - V_{b,2} \\ V_{a,2} - V_{b,3} \\ V_{a,3} - V_{b,1} \end{bmatrix}, \qquad (10)$$

Where $a$ and $b$ represent the top and bottom nodes, respectively. $I_{a,i}\ (i=1,2,3)$ represents the input current at node $i$. $V_{a,i}\ (i=1,2,3)$ and $V_{b,i}\ (i=1,2,3)$ represent the voltages at the top and bottom node $i$, respectively. Assuming no external sources, by diagonalizing this equation using a unitary transformation $F = \frac{1}{\sqrt{3}} \begin{bmatrix} 1 & 1 & 1 \\ 1 & e^{i2\pi/3} & e^{i4\pi/3} \\ 1 & e^{i4\pi/3} & e^{i8\pi/3} \end{bmatrix}$, we obtain:

$$-\omega^2 C \begin{bmatrix} 0 & 0 & 0 \\ 0 & 3 & 0 \\ 0 & 0 & 3 \end{bmatrix} \begin{bmatrix} V_{a,1}^* \\ V_{a,2}^* \\ V_{a,3}^* \end{bmatrix} = (\omega^2 C_g - \frac{i}{\omega L}) \begin{bmatrix} V_{a,1}^* \\ V_{a,2}^* \\ V_{a,3}^* \end{bmatrix} + \frac{i}{\omega L} \begin{bmatrix} 1 & 0 & 0 \\ 0 & e^{i2\pi/3} & 0 \\ 0 & 0 & e^{-i2\pi/3} \end{bmatrix} \begin{bmatrix} V_{b,1}^* \\ V_{b,2}^* \\ V_{b,3}^* \end{bmatrix}, \qquad (11)$$

where $V_a^* = F[V_{a,1}, V_{a,2}, V_{a,3}]^T$ represents the transformed voltage basis. Notably, one of these basis components $V_{a,1}^* = V_{a,1} + V_{a,2} + V_{a,3}$ is independent of $\omega$, leaving $V_{a,2}^*$ and $V_{a,3}^*$ as a pair of decoupled voltage pseudospins $V_{a,\uparrow}=V_{a,1}+e^{i4\pi/3}V_{a,2}+e^{i2\pi/3}V_{a,3}$ and $V_{a,\downarrow}=V_{a,1}+e^{i2\pi/3}V_{a,2}+e^{i4\pi/3}V_{a,3}$. These two voltage pseudospins satisfy the following independent equations:

$$EV_{a,\uparrow}=UV_{a,\uparrow}+\frac{i}{\omega L}e^{i2\pi/3}V_{b,\uparrow},$$

$$EV_{a,\downarrow}=UV_{a,\downarrow}+\frac{i}{\omega L}e^{-i2\pi/3}V_{b,\downarrow}. \quad (12)$$

where we have $E=-3\omega^2 C$ and $U=\omega^2 C_g - i/\omega L$. It is clearly shown that the effective coupling with a phase factor being $\varphi=\pm 2\pi/3$ can be realized in the subspace of each voltage pseudospin.

Furthermore, the hopping terms with other phase factors $\frac{2\pi}{N}\cdot i\,(i=1,2,\ldots,N)$ can also be constructed by generalizing $N$ subnodes representing a single node following the same method [16].

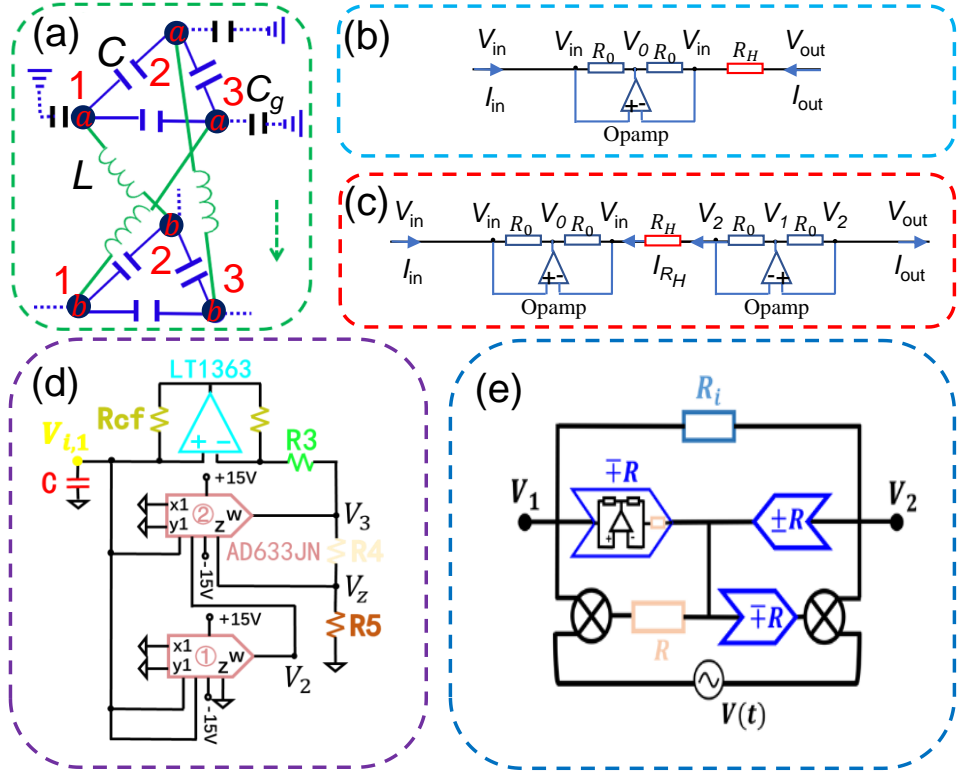

**Figure 1**. (a) The circuit diagram for hopping term $e^{i2\pi/3}$. (b) Non-reciprocal resistance circuit. (c) Negative resistance circuit. (d) Nonlinear effects in circuits. (e) Time-varying couplings in circuits.

**2.2.2 Non-Reciprocal and negative couplings in circuits**

In lattice models, non-reciprocal hopping terms can be realized in circuits using a negative impedance converter through current inversion (INIC) [25, 26]. This setup allows for asymmetric hopping, where hopping in one direction is negative, while in the opposite direction, it is positive. Figure 1(b) illustrates a non-reciprocal resistance circuit. The key component in this design is the operational amplifier (op-amp), which, in a negative feedback configuration, ensures that the voltages at the negative input terminals of the op-amp equals to the voltages at the positive input terminals $V_{in}$. Since the currents flowing into the inputs of the op-amp are very small, approximately zero, the currents $I_{in}$ and $I_{out}$ can be expressed as:

$$\begin{pmatrix} I_{in} \\ I_{out} \end{pmatrix} = \frac{1}{R_0} \begin{pmatrix} -1 & 1 \\ -1 & 1 \end{pmatrix} \begin{pmatrix} V_{in} \\ V_{out} \end{pmatrix}. \tag{13}$$

It can be seen that $I_{in}$ equals $I_{out}$ instead of $-I_{out}$. In this case, the reciprocity is broken, and this setup effectively simulates non-reciprocal hopping term.

Based on two op-amps, we can also realize the negative resistance coupling, being functionalized as a negative resistance. Figure 1(c) illustrates a negative resistance circuit. According to Equation (13), the currents $I_{in}$ can be expressed as:

$$\begin{pmatrix} I_{in} \\ I_{R_H} \end{pmatrix} = \frac{1}{R_0} \begin{pmatrix} -1 & 1 \\ -1 & 1 \end{pmatrix} \begin{pmatrix} V_{in} \\ V_{out} \end{pmatrix}. \tag{14}$$

Besides, the currents $I_{out}$ and $I_{R_H}$ can be expressed as:

$$I_{out} = \frac{1}{R_0}(V_1 - V_{out}),$$

$$I_{R_H} = \frac{1}{R_0}(V_1 - V_{out}). \tag{15}$$

Combining Equations (14) and (15), the matrix formalism of $I_{in}$ and $I_{out}$ can be expressed as

$$\begin{pmatrix} I_{in} \\ I_{out} \end{pmatrix} = \frac{1}{R_0} \begin{pmatrix} -1 & 1 \\ -1 & 1 \end{pmatrix} \begin{pmatrix} V_{in} \\ V_{out} \end{pmatrix}. \tag{16}$$

Although Equation (16) is the same as Equation (14), the direction of $I_{out}$ in Figures 1(b) and (c) are different. So, Equation (16) indicate the direction of currents $I_{in}$ and $I_{out}$ is different from the direction of decreasing voltage. This is the opposite of the current-voltage relationship of a normal resistance. It effectively simulates a negative resistance.

### 2.2.3 Nonlinear Stuart-Landau oscillators in circuits

In this section, we provide details on the correspondence between Chua diode and nonlinear Stuart-Landau oscillator [27]. Figure 1(d) presents the schematic diagram of Chua diode, which is composed by two multipliers and an INIC. Based on the relationship between the input and output voltages of the first and second multipliers, we can obtain two relationships of

$$V_2 = \frac{V_{i,1}^2}{10}, V_3 = \frac{V_{i,1}*V_2}{10}\frac{R_4+R_5}{R_4}. \tag{17}$$

Combing these two terms, it have $V_3 = \frac{R_4+R_5}{100R_4}V_{i,1}^3$. By applying Kirchhoff's law on the circuit node of $V_{i,1}$, we obtain the following equation as

$$\frac{dV_{i,1}}{dt} = \frac{1}{CR_3}V_{i,1} - \frac{R_4+R_5}{100CR_3R_4}V_{i,1}^3. \tag{18}$$

Equation (18) possesses the same form to the dynamical equation of Stuart-Landau oscillator with effective parameters being $\alpha = \frac{1}{CR_3}$ and $\beta = \frac{R_4+R_5}{100CR_3R_4}$.

### 2.2.4 Time-varying couplings in circuits

Lastly, we provide the detailed derivation to illustrate the realization of time-varying INIC [28], as shown in Figure 1(e). There are three conventional INICs $\pm R$, two multipliers and two resistances $R$ and $R_i$ in a single time-varying INIC. The transfer function of the multiplier is $V_{out}(t) = V_{in1}(t)*V_{in2}(t)/10$, where $V_{out}(t)$ is the output-node voltage of the multiplier and $V_{in1}(t)$ and $V_{in2}(t)$ are two input-node voltages of the multiplier. To realize time-modulated INIC, the external voltage $V(t)$ is injected into one input node of the multiplier. Carrying out the Kirchhoff's law on three circuit nodes labeled by $V_1$, $V_0$ and $V_2$, we obtain three following equations as

$$I_1 = \frac{V_1-V_0}{-R} + \frac{V_1-V_2}{R_i}, \tag{19}$$

$$I_2 = \frac{V_2-V_0}{-R} + \frac{V_2-V_1}{R_i}. \tag{20}$$

$$I_0 = \frac{V_0 - V_1}{R} + \frac{V_0 - V_2}{R_i} + \frac{V_0 - \frac{V_1 V(t)}{10}}{R} + \frac{V_0 - \frac{V_2 V(t)}{10}}{-R}, \quad (21)$$

where $I_1$, $I_0$ and $I_2$ correspond to currents flowing into three circuit nodes. By setting $I_1$, $I_0$ and $I_2$ to zero, and combing Equations (19)-(21), we can express $V_0$ as

$$V_0 = \frac{1}{2}(V_1 + V_2) + \frac{V(t)}{20}(V_1 - V_2). \quad (22)$$

Then, substituting Equation (22) into Equations (19) and (20), we have

$$I_1 = V_1\left(\frac{1}{R_i} - \frac{1}{2R} + \frac{V(t)}{20R}\right) + V_2\left(-\frac{1}{R_i} + \frac{1}{2R} - \frac{V(t)}{20R}\right), \quad (23)$$

$$I_2 = V_2\left(\frac{1}{R_i} - \frac{1}{2R} - \frac{V(t)}{20R}\right) + V_1\left(-\frac{1}{R_i} + \frac{1}{2R} + \frac{V(t)}{20R}\right). \quad (24)$$

When $R_i = 2R$, Equations (23)-(24) can be written as

$$\begin{pmatrix} I_1 \\ I_2 \end{pmatrix} = \frac{V(t)}{20R}\begin{pmatrix} 1 & -1 \\ 1 & -1 \end{pmatrix}\begin{pmatrix} V_1 \\ V_2 \end{pmatrix}. \quad (25)$$

It is shown that the currents do not fulfil the reciprocity condition of $I_1 = -I_2$ as for a passive circuit element. In this case, our designed circuit element acts as a time-varying INIC, where the current flowing into the element at the left is different from the current coming out at the right.

### 2.3. Experimental techniques for characterizing circuit networks

When analogizing the time-dependent Schrodinger equation, simultaneous evolution of signals at different nodes of circuit networks is required. This can be achieved by relays which can disconnect the signals simultaneously. The time-varying output voltage can be measured using an oscilloscope. There are two cases when analogizing the stable-state Schrodinger equation. One case is that the voltages of circuit can be written as Equation (8). In such case, the $\beta(\omega)$ and the voltages $V_i$ $(i = 1, 2, \ldots N)$ correspond to the eigenvalues and eigenstates of the lattice model. In experiment, the $\beta(\omega)$ and the voltages $V_i$ $(i = 1, 2, \ldots N)$ can be measured directly with the impedance analyzer and oscilloscope.

Another case is that the voltages of circuit can be written as Equation (9). In such case,

only at a certain frequency $\omega_0$, the circuit Laplacian $J(\omega_0)$ corresponds to the Hamiltonian *H*. The properties of quantum systems can be analyzed by reconstructing the admittance spectrum using voltage measurement and Fourier transform [29, 30]. When left-multiplying both sides of Equation (7) by $G(\omega)$ which is the inverse matrix of $J(\omega)$, expression $G(\omega)I(\omega)=V(\omega)$ is obtained. Here $G(\omega)$ is called the impedance matrix. When exciting only one node of the circuit such as node *j* at a certain frequency $\omega_0$, and then measuring the input current $I_j$ and the voltages $V_i(i=1,2,\ldots N)$ of all *N* nodes. One column of the impedance matrix $G(\omega_0)$ can be obtained by $G_{ij}=V_i/I_j$. After exciting the remaining nodes in the circuit and repeating the measurement of the input current and voltages for *N* nodes, it is possible to obtain the whole impedance matrix $G(\omega_0)$, which can then be inverted to acquire the circuit Laplacian $J(\omega_0)$.

## 3. Exploring topological states of matter based on electric circuits

The study of topological states of matter has seen rapid advancements in recent years, becoming a central focus in fields such as condensed matter physics and metamaterials. Pioneering research into topological states has provided essential insights into quantum phase transitions and the emergence of novel phases of matter in Hermitian systems, deepening our understanding of the fundamental properties of quantum materials. However, many real-world systems exhibit non-Hermitian characteristics, where energy is not conserved, revealing a host of intriguing topological phenomena that were previously inaccessible. This realization has sparked considerable interest in exploring topological states within non-Hermitian frameworks, opening up exciting new avenues for both theoretical and experimental research. In addition, the introduction of nonlinearities, time-varying modulations, and few-body correlation effects into topological systems has the potential to generate a wide range of novel and exotic topological states. These developments could lead to the discovery of entirely new classes of topological phases, further enriching our understanding of quantum materials and paving the way for innovative technological applications.

Topolectrical circuits have emerged as a particularly promising tool for probing these diverse topological states. Through the careful design of circuit structures, researchers can simulate and manipulate various topological states, making topolectrical circuits a versatile and accessible platform for experimentation. Compared to traditional quantum simulation methods, topolectrical circuits offer several advantages, including simplicity in design, precise controllability of parameters, and ease of experimental verification. In this section, we examine recent progress in using topolectrical circuits to simulate various types of topological states. We begin by discussing advancements in the exploration of Hermitian topological states through electrical circuits, progressing from first-order to higher-order topologies, and within each section, from lower to higher geometric dimensions. Following this, we delve into the exploration of topological states in non-Hermitian circuit networks. The review then summarizes recent developments in nonlinear and time-varying topolectrical circuits, highlighting their potential to realize new topological phenomena. Finally, we explore the use of topolectrical circuits in simulating few-body interaction systems.

## 3.1 Engineering Hermitian topological states in electric circuit networks

The realization of Hermitian topological states within electric circuits has provided a new framework for understanding topological phases. In this part, we organize the discussion by moving from first-order to higher-order topologies and within each section, presenting topics according to increasing geometric dimensions.

### 3.1.1 First-order topological states in Hermitian electric circuits

A defining feature of any topological phase of matter is the bulk-boundary correspondence, which guarantees the existence of topologically protected boundary modes that are robust against perturbations. Typically, a $d$-dimensional topological phase hosts these modes on ($d-1$)-dimensional boundaries, characterized by a co-dimension of one, corresponding to the first-order topological phases. Here, we summarize the realization of several first-order topological insulating phases within topolectrical circuits.

**3.1.1.1. Topolectrical circuits for several first-order topological insulating models.**

The Su-Shrieffer-Heeger (SSH) model is a quintessential example of a topological state in one dimension [31]. In this section, we describe its realization in electrical circuits [17, 20, 32]. Figure 2a illustrates the schematic diagram of the 1D SSH circuit, which consists of a linear array of inductors and capacitors arranged in a repeating pattern. Each unit cell of the circuit contains two nodes connected by alternating intra-cell and inter-cell capacitors, $C_1$ and $C_2$, respectively. The inductors are connected between the nodes and ground, providing the necessary reactive components to form the desired circuit topology. The key feature of this circuit is its admittance matrix, which governs the relationship between the currents and voltages at each node. The corresponding circuit Laplacian $J_{SSH}(k_x)$ for the SSH circuit can be expressed as:

$$J_{SSH}(k_x) = i\omega \left( C_1 + C_2 - \frac{1}{\omega^2 L} \right) \mathbf{I} \\ -i\omega \left[ (C_1 + C_2 \cos k_x) \sigma_x + C_2 \sin k_x \sigma_y \right] \quad (34)$$

where $k_x$ is the wavevector and $\omega$ is the angular frequency. This matrix closely mirrors the Hamiltonian of the 1D SSH model, where the intra- and inter-cell hopping amplitudes are determined by the capacitive couplings $C_1$ and $C_2$. The topological properties of the 1D SSH model, such as the existence of midgap zero-energy edge states, can therefore be simulated by tuning the ratio of $C_1$ to $C_2$. The clear correspondence between the circuit Laplacian and the Hamiltonian of the 1D SSH model demonstrates how electrical circuits can effectively simulate and probe the topological properties. In this case, to characterize the topological properties of the SSH topolectrical circuit, impedance measurements are employed, as shown in Figure 2b. By measuring the impedance between the nodes at the boundaries of the circuit, it is possible to detect the presence of topologically protected boundary modes. Specifically, when $C_1 < C_2$ (black line), the impedance peak can locate at the frequency matched to zero eigen-energy, corresponding to the existence of edge states in the topologically nontrivial phase. Conversely, when $C_1 > C_2$ (red line), there is no high impedance peak at zero energy, indicating the absence of edge states and a transition to the trivial phase. These impedance results provide a direct way to observe the topological phase transition in the SSH model realized within topolectrical circuits. In addition, Li et al have constructed a topological inductor–capacitor circuit, which consists of two identical SSH chains coupled using a middle chain with capacitors, to observe topological bound states in the continuum [33]. It may be suitable for electrical device applications, such as antennae, filters, and radio frequency devices.

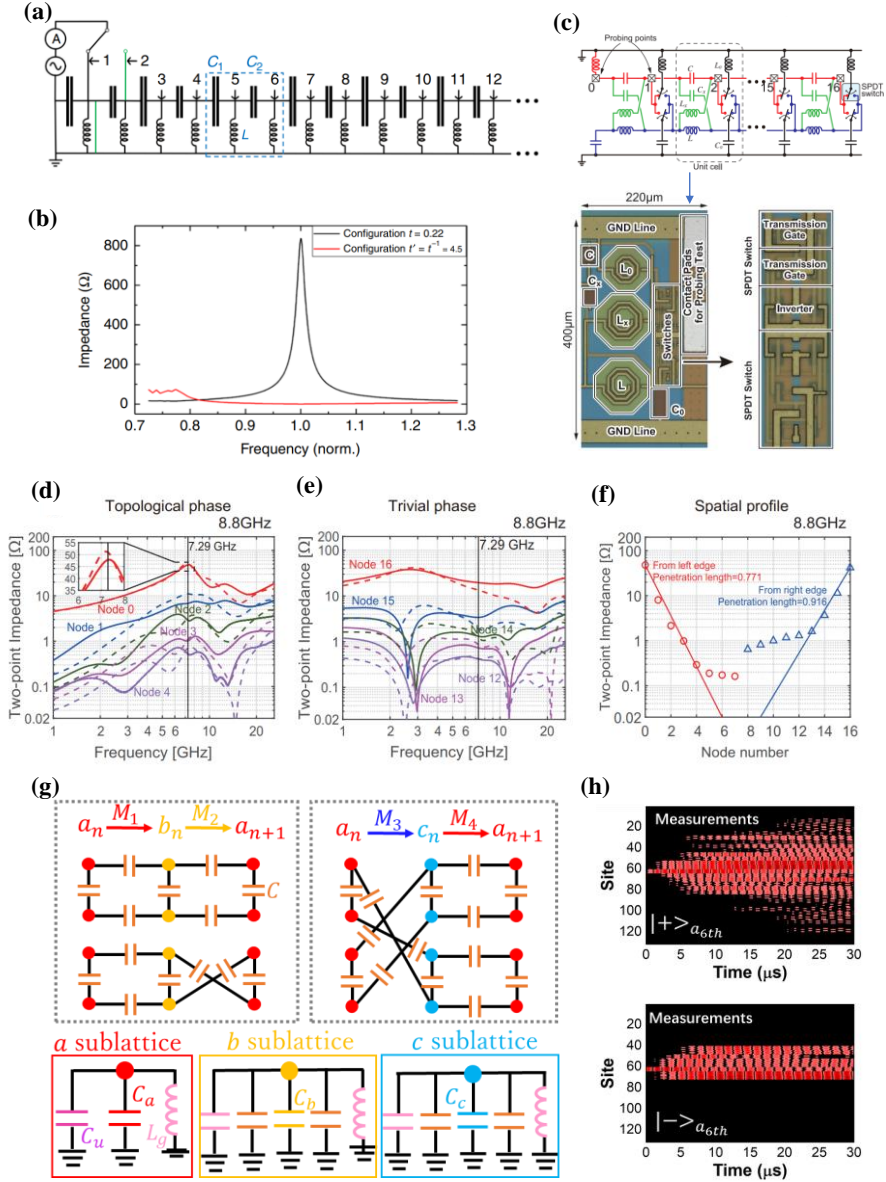

**Figure 2. 1D SSH and Kitaev topolectrical circuits.** (a). The schematic diagram of the 1D SSH circuit. (b). Experimental results of impedance measurements for 1D non-trivial SSH circuit. Figures a and b are taken from [17]. Figures are adapted from CC-BY open access publications Springer Nature. (c). The schematic diagram of the 1D Kitaev circuit. (d) and (e). Experimental results of impedance measurements for 1D topological and trivial Kitaev circuits. (f). The measured spatial profile of the impedance values for 1D topological Kitaev circuits. Figures c-f are taken from [35]. Figures are adapted from CC-BY open access publications

Springer Nature. (g). The schematic diagram of 1D non-Abelian AB-cage topolectrical circuit. (h). Experimental results of voltage dynamics with non-Abelian inverse Anderson transition. Figures g and h are taken from [38]. Reproduced with permission [38] Copyright 2023, American Physical Society.

Except for the 1D SSH model, the Kitaev topological superconductor model is an another intriguing 1D systems realizing topological insulators and superconductors. Ezawa et al have proposed a theoretical proposal with the use of chains of capacitors and inductors [34] to simulate Majorana-like edge states. Recently, Iizuka et al have designed and fabricated a switchable structure in the integrated circuit to control the position of a Majorana-like interface state arbitrarily along a chain [35], as shown in Figure 2c. By measuring impedance spectra of topological and trivial Kitaev integrated circuits, as shown in Figures 2d and 2e, midgap edge states were observed. Figure 2f presents the spatial profile of the impedance values for all-topological mode measured from both left and right edges, being consistent to the distribution of topological edge states with strong boundary localization.

In the last part of this section, we summarize the recent development on the implementation of 1D Aharonov-Bohm (AB) cages by topolectrical circuits. AB cage is a fascinating phenomenon in condensed matter physics, where magnetic flux threading through specific lattice structures leads to the complete localization of wave functions. This localization occurs due to destructive interference, preventing electrons from propagating through the lattice. The significance of AB cages lies in their ability to induce the formation of topologically flat bands. These flat bands are of particular interest because they can enhance electronic correlations and give rise to exotic quantum states. It has been demonstrated that the introduction of disorder into a system with AB cages can lead to a remarkable transition known as the inverse Anderson transition [36]. Unlike the traditional Anderson localization, where disorder in a system causes a transition from a metallic (delocalized) state to an insulating (localized) state, the inverse Anderson transition describes the opposite behavior. Here, the system transitions from an AB-cage localized state to a delocalized state, effectively reversing the localization induced by the AB cages. Although the inverse Anderson transition has already been proposed, the experimental realization of such a phenomenon is not an easy task. Wang et al have reported the implementation of inverse Anderson transitions based on AB-cage topolectrical circuits with symmetric-correlated, antisymmetric-correlated, and uncorrelated disorders [37]. Through the direct measurements of frequency-dependent impedance responses and time-domain voltage dynamics, the inverse Anderson transitions induced by antisymmetric-correlated disorders are clearly observed. Furthermore, Zhang et al have further extended the 1D AB-cage topolectrical circuit to the non-Abelian case through the construction of nilpotent interference matrix (Figure. 2g), and observed non-Abelian inverse Anderson transitions [38]. It is found that pseudospin-dependent localized and delocalized eigenstates coexist in the disordered non-Abelian AB-cage topolectrical circuit, making inverse Anderson transitions depend on the relative phase of two internal pseudospins (Figure. 2h). These disordered AB-cage topological circuits establish a new platform for investigating the interplay between disorder and topological flat bands. In addition, except for the inverse Anderson transition, Wang et al have also reported on the experimental realization of non-Abelian Anderson localization and transition based on engineered topolectrical circuits, which are directly mapped to the quasiperiodic Aubry-André-Harper model with non-Abelian gauge fields [39].

**3.1.1.2. 2D Chern topolectrical circuits and time-reversal invariant topolectrical circuits**

Chern insulators with time-reversal symmetry and time-reversal invariant topological insulators are two representative topological states in two dimensions, characterized by their robust edge modes and topological invariants. By engineering the circuit connection and symmetry, these 2D topological states can be constructed in circuit networks. Hofmann et al. theoretically designed a 2D circuit network that breaks time-reversal symmetry to implement a circuit analog of a Chern insulator, featuring unidirectional voltage modes [40]. The schematic diagram of this 2D circuit is presented in Figure 3a. The circuit is constructed from a network of INICs, inductors, and capacitors arranged in a square lattice, where the non-reciprocal property of INICs breaks the time-reversal symmetry of the circuit. The circuit eigen-equations can be derived from Kirchhoff's laws, yielding a circuit Laplacian $J_{Chern}(k_x, k_y)$, that closely parallels the Hamiltonian of a Chern insulator:

$$J_{TTC}(\vec{k};\omega) = i\omega\left(3C_0 + C_g - \frac{1}{\omega^2 L_0}\right)\mathbf{I} - i\omega C_0\left[1 + \cos(k_x) + \cos(k_y)\right]\sigma_x$$
$$-i\omega C_0\left[\sin(k_x) + \sin(k_y)\right]\sigma_y + i\omega\left[\Delta + \frac{2}{\omega R_0}\left[\sin(k_x) - \sin(k_y) - \sin(k_x - k_y)\right]\right]\sigma_z.$$

(35)

The topological nature of this circuit is manifested in the presence of unidirectional edge states for voltages. These edge states are detectable through time-domain voltage measurements at the circuit boundaries, showing the characteristic defect-free one-way transmission that is the hallmark of Chern topological insulators, as shown in Figure 3b. There are also several works on the realization of Chern-class insulators by electric circuits, such as Chern circuits with arbitrarily large Chern numbers [41].

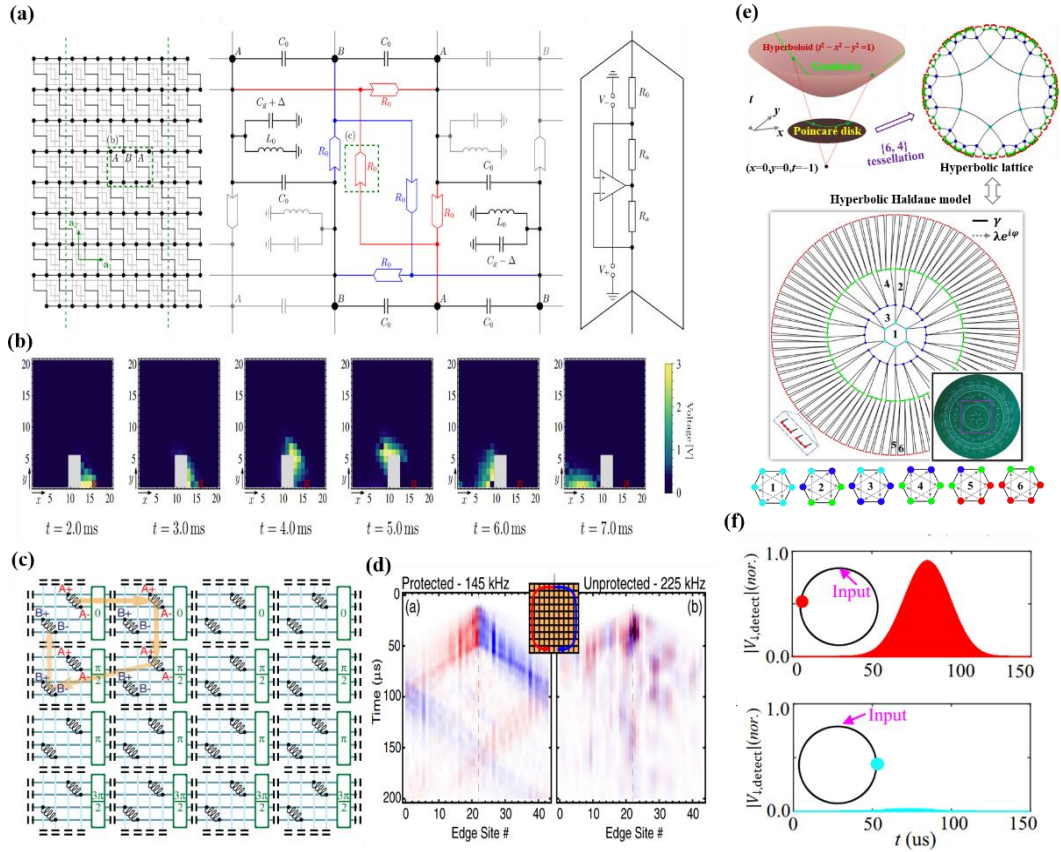

**Figure 3. 2D Chern and time-reversal invariant topolectrical circuits in Euclidean and hyperbolic lattices.** (a). The schematic diagram of the 2D Chern circuit without time-reversal symmetry. (b). Simulation results of voltage evolution in the form of the defect-free one-way transmission along the boundary of 2D Chern circuit. Figures a and b are taken from [40]. Reproduced with permission [40] Copyright 2019, American Physical Society. (c). Time-

reversal invariant capacitor-inductor topological circuit network with voltage pseudospins-dependent effective magnetic fields. (d). Experimental results for the spin-momentum locked one-way propagation along the boundary of time-reversal invariant topological circuit network. Figures c and d are taken from [15]. Figures are adapted from CC-BY open access publications American Physical Society. (e). The hyperbolic Haldane model with the photon of the hyperbolic circuit sample shown in the inset. (f). Experimental measurements of voltage evolution in hyperbolic topological circuits. Figures e and f are taken from [63]. Figures are adapted from CC-BY open access publications Springer Nature.

Then, we summarize the circuit implementation of 2D topological states with the spin degree of freedom, that preserve time-reversal symmetry. Jia et al have demonstrated the first time-reversal invariant capacitor-inductor topological circuit network [15], as shown in Figure 3c. The effective pseudospins can be realized by introducing an internal degree of freedom into each effective lattice site (four circuit nodes acting as a lattice site) of the circuit. In this case, by engineering the complex coupling phases between voltage pseudospins at nearby effective lattice sites, the effective magnetic fields can be introduced into the circuit, giving rise to modified Hofstadter spectrum. Figure 3d displays time-resolved transport dynamics of the topological edge modes in circuits, showing the spin-momentum locked one-way voltage propagation along the boundary. In the same year, a general theory on the realization of voltage pseudospins with fewer elements per site has been proposed by Albert et al [16], where several values of complex coupling phases can be realized based on this method. Lately, by introducing an internal degree of freedom, Zhu et al. have designed a circuit model that realizes quantum spin-valley locked topological circuits [42, 43], being analog to Kane-Mele Hamiltonian [44]. The key to achieving the spin-valley Hall effect in this setup is the coupling phase between the networks, which can be tuned to simulate the spin-orbit interaction found in quantum spin Hall insulators. For this purpose, four circuit nodes are coupled to construct a pair of voltage pseudospins. By suitably exciting circuits with a fixed voltage pseudospin, the voltage evolution in the circuit reveals the spin-selective unidirectional transport properties, indicative of the spin-polarized edge states associated. Following the same method on the construction of voltage pseudospins, Yang et al have constructed the electric circuits with pseudospins, which are mapped to a modified Haldane model with anti-chiral edge states in each pseudospin subspace [45].

There are also various studies on the demonstration of 2D topological states by electric circuits, including the circuit implementation of a 2D weak topological insulator [46], Hopf insulator [47], 2D moiré circuits [48], 2D non-Abelian topological circuits exhibiting the effective spin-orbit interaction and topological Chern states driven by non-Abelian gauge fields [49] and others [50-53]. In the microwave domain, Inductance-Capacitance (LC) circuits with planar microstrip arrays have been meticulously designed and fabricated, enabling the realization of topological edge states that selectively respond to orbital angular momentum [54-60]. These breakthroughs highlight the growing capabilities of topolectrical circuits in manipulating complex topological phenomena.

Beyond the widely revealed topological states in 2D Euclidean lattices, it is worth noting that exotic 2D topological states also exist in hyperbolic lattices [61-69], which are regular tessellations in non-Euclidean space with constant negative curvature. These lattices support novel topological phases that do not have direct analogs in Euclidean geometries. Zhang et al. have extended the Haldane model, a prototypical model for Chern insulators, to hyperbolic lattices (Figure 3e), designing a 2D topolectrical circuit to realize hyperbolic topological states with significantly enhanced boundary responses [63]. The right-bottom inset of Figure 2f shows the photograph imagine of the circuit sample. The circuit is structured on a hyperbolic lattice, where the connectivity of the nodes follows the hyperbolic tessellation rules. It is noted that to introduce complex hoppings in hyperbolic circuits, three circuit nodes are suitably connected to construct voltage pseudospins exhibiting $e^{\pm i2\pi/3}$ coupling phases. Thus, derived from Kirchhoff's laws, the circuit Laplacian with respect to a voltage pseudospin is consistent with the Hamiltonian of the hyperbolic Haldane model. The effective Hamiltonians of two effective voltage pseudospin are time-reversal counterparts with each other. The consistency between the circuit equations and the quantum model's equations allows for the direct simulation of hyperbolic topological states. As shown in Figure 3f, the experimental measurements of voltage evolution reveal the realization of boundary-dominated hyperbolic boundary modes, demonstrating the robustness and richness of topological states in non-Euclidean geometries. Beyond the low-frequency domain in electric circuits, recent investigations have demonstrated the realization of 2D hyperbolic topological phases in microwave networks [70], and coupled ring resonators in the near-infrared region [71], giving the foundation for designing topological devices with high-efficient structure utilization.

**3.1.1.3. 3D topological semimetals in topolectrical circuits**

Except for the gapped topological insulators, topological band theory also encompasses gapless topological phases, known as topological semimetals (TSMs) [72, 73], which feature nodal points or lines in their band structures protected by topology. To date, several types of TSMs have been implemented in 3D topolectrical circuits. Figure 4a presents the schematic diagram of a 3D topolectrical circuit designed by Luo et al. to simulate a nodal-line semimetal [74]. This circuit consists of a 3D network of inductors and capacitors, where each unit cell contains nodes connected by capacitors and inductors, arranged to create a nodal line in the band structure. The presence of the nodal line in the admittance spectrum corresponds to the gapless topological feature of a type-B nodal-line semimetal, as shown in Figure 4b. Additionally, another typical TSMs, Weyl semimetals, has also been realized in 3D topolectrical circuits [75]. In this setup, nontrivial winding in the Weyl points' band structure leads to the emergence of Fermi arcs at the circuit boundaries. These features are observable through the unique boundary impedance responses. Moreover, in the fields of Weyl topolectrical circuits, Li et al. have reported the realization of ideal type-II Weyl TSMs by stacking LC resonator dimers with broken parity inversion symmetry [76], as shown in Figure 4c. Further extending topolectrical circuits to quantum anomalous semimetals, which correspond to Wilson fermions, have also been implemented [77]. Figure 4d presents the schematic diagram of such a circuit, where the circuit's nontrivial topology is characterized by the existence of chiral edge currents. The chiral edge currents are observed by forming a domain wall separating two circuits with opposite fractional Chern numbers. Except for the above TSMs with simple geometries of nodal points or nodal lines, Lee et al. have designed and fabricated a Resistance-Inductance-Capacitance (RLC) circuit network, where the constructed nodal lines are in the form of different types of knots in momentum space [78]. Rafi-Ul-Islam et. al have realized the Chiral surface and hinge states in higher-order Weyl semimetallic circuits [79].

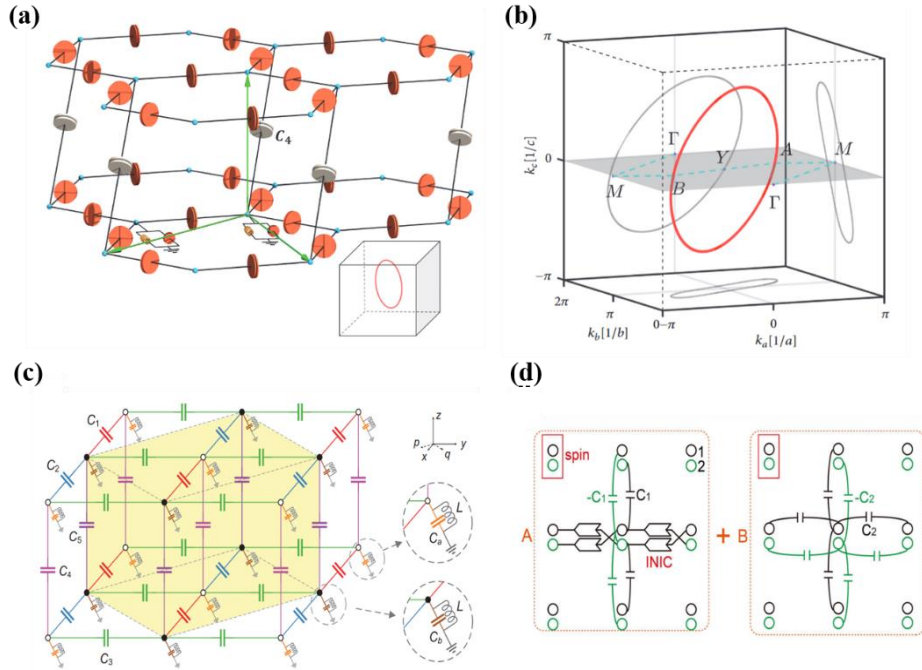

**Figure 4. 3D topological semimetals in topolectrical circuits.** (a). The schematic diagram of a 3D topolectrical circuit simulating a type-B nodal-line semimetal. (b). Simulation results of the nodal line in the admittance spectrum. Figures a and b are taken from [74]. Figures are adapted from CC-BY open access publications AAAS. (c). The schematic diagram of type-II Weyl circuit. Figure c is taken from [76]. Figure is adapted from CC-BY open access publications Oxford University Press. (d). The schematic diagram of a circuit simulating Wilson fermions. Figure d is taken from [77]. Figure is adapted from CC-BY open access publications Springer Nature.

### 3.1.1.4. 4D topological insulators and 5D Weyl physics in topolectrical circuits

Circuit-based systems are particularly well-suited for exploring high-dimensional topological states due to their design flexibility, controllability, and ease of experimental implementation. A groundbreaking realization of a 4D topological insulator was achieved using electrical circuits, where a 4D lattice of interconnected components was engineered to reflect the topological characteristics of a 4D system. Yu et al have reported the first theoretical design of circuit networks to realization 4D spinless topological insulator in class AI [80], as shown in Figure 5a. It is demonstrated that the frequency spectrum of designed circuits exhibits pairs of 3D Weyl boundary states, a hallmark of the nontrivial topology, as shown in Figure 5b. Lately, Wang designed and fabricated a circuit network to experimentally explore 4D topological insulators in AI class [81]. Figure 5c illustrates the schematic diagram of the designed 4D topolectrical circuit, where the circuit nodes are interconnected in a manner that simulates a 4D lattice structure. The circuit equations yield an admittance matrix that parallels the Hamiltonian of a 4D topological insulator. This implementation allowed for the simulation of topologically protected edge states at the 4D circuit boundaries, with the presence of these states observable through distinct impedance signatures, as shown in Figure 5d. In addition, a 4D topological circuit has been designed by Li et al. to realize a full 3D-imaging of nodal boundary Seifert surfaces [82]. Extending this capability further, 5D Weyl physics has been realized in electrical circuits [83], allowing researchers to explore the properties of Yang monopoles and Weyl surfaces in five dimensions. Figure 5e shows the schematic diagram of a 5D topolectrical circuit designed to simulate these high-dimensional topological features. The circuit equations yield an admittance matrix that mirrors the Hamiltonian of a 5D Weyl semimetal, where the Yang monopoles and Weyl surfaces are observable by recovering the admittance matrix.

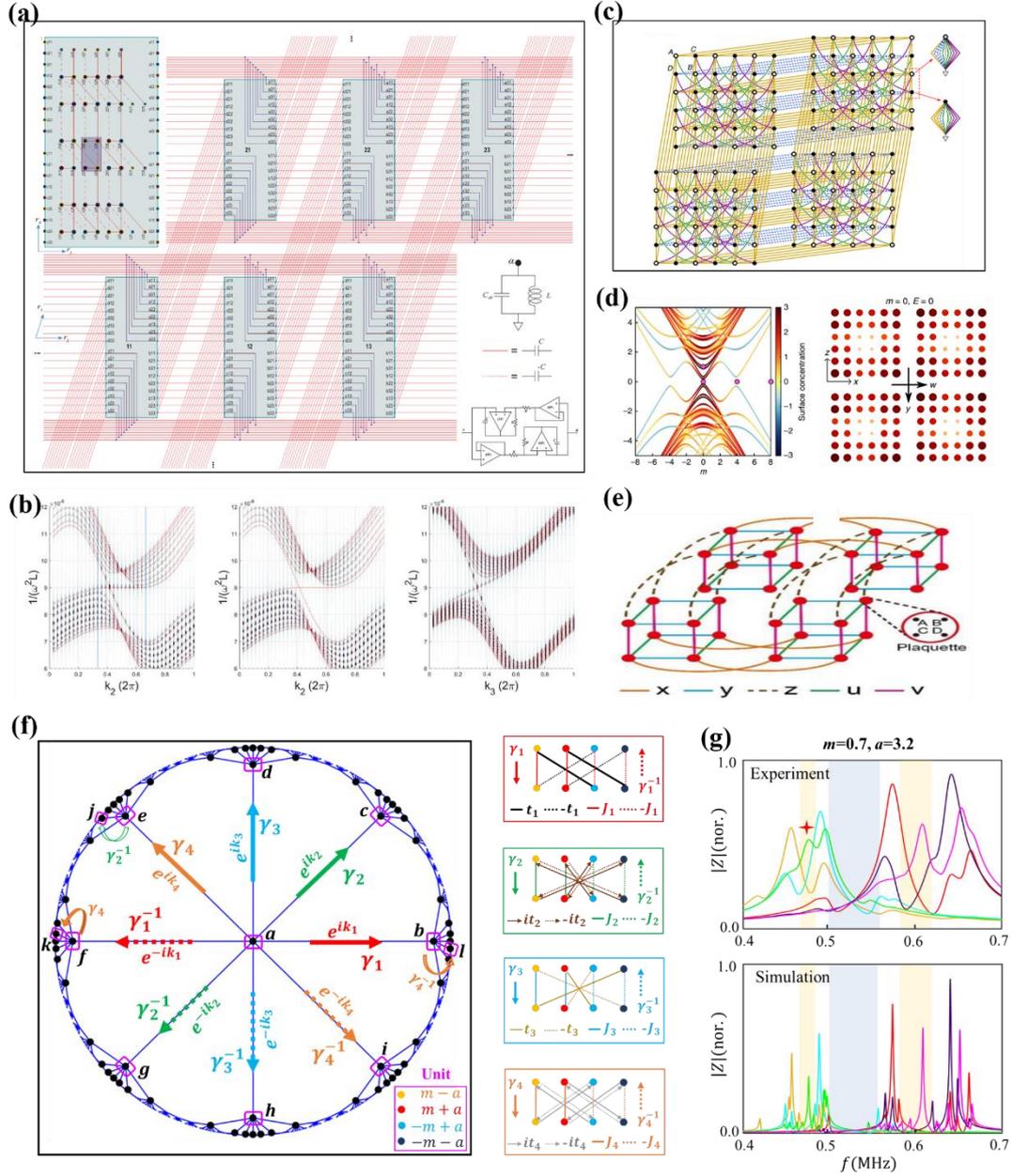

**Figure 5. 4D topological insulators and 5D Weyl physics in topolectrical circuits.** (a). The schematic diagram of circuit networks for realizing 4D spinless topological insulator in class AI. (b). Simulated 3D Weyl boundary states of 4D spinless topological circuit in class AI with open boundaries along different directions. Figures a and b are taken from [80]. Figures are adapted from CC-BY open access publications Oxford University Press. (c). The schematic diagram of the designed 4D topolectrical circuit with circuit nodes being interconnected in a manner that simulates a 4D lattice structure. (d). Results of band structures and spatial profiles of surface states in 4D topolectrical circuits with open boundaries. Figures c and d are taken from [81]. Figures are adapted from CC-BY open access publications Springer Nature. (e). The schematic diagram of a 5D topolectrical circuit with Yang-monopoles and Weyl surfaces. Figure e is taken from [83]. Figure is adapted from CC-BY open access publications American Physical Society. (f). The schematic diagram of topological insulator in {8, 8} hyperbolic Bravais lattice. (g). Measured impedance spectra of hyperbolic topological circuits. Figures f and g are taken from [86]. Figures are adapted from CC-BY open access publications Springer Nature.

Furthermore, the recent ground-breaking implementations of hyperbolic lattices have stimulated numerous advances in hyperbolic physics. The advent of hyperbolic band theory has opened new possibilities for constructing topological band theories in non-Euclidean spaces [84, 85]. It is noted that several hyperbolic lattices exhibit Fuchsian translational symmetry groups, which is a non-Abelian translational group and features both one-dimensional and higher-dimensional group representations. Interestingly, it has been found that due to the non-Abelian nature of the Fuchsian translational group, 2D hyperbolic lattices can possess high-dimensional momentum spaces. Motivated by this novel property, Zhang et al. have constructed a topological lattice model with an {8, 8} hyperbolic Bravais lattice that features an effective four-dimensional momentum space [86], as shown in Figure 5f. By tuning the coupling parameters, a hyperbolic topological insulator protected by the second Chern number defined in 4D $k$-space has been constructed. Furthermore, they designed and fabricated a hyperbolic circuit network and experimentally observed hyperbolic topological boundary states induced by the second Chern number, as shown in Figure 5g. This work reveals novel hyperbolic topological states with the second Chern number defined in four-dimensional momentum space within two-dimensional hyperbolic lattices. Unlike previous experiments that achieved four-dimensional quantum Hall effects through two-dimensional topological pumping and non-local circuit connections, this work's momentum space dimension extension is entirely due to the non-Abelian characteristics of the Fuchsian translational symmetry group in negative curvature space.

### 3.1.2. Higher-order topological states in Hermitian electric circuits

The concept of bulk-boundary correspondence, which has been central to understanding first-order topological phases, has been extended to account for low-dimensional topological modes residing on boundaries with an integer codimension larger than one. This extension leads to what are known as higher-order topological (HOT) phases [87, 88]. Notable examples of HOT phases include the corner modes in 2D systems and hinge modes in 3D systems. In this section, we summarize the exploration of HOT insulators in electrical circuits, covering examples from 2D to 4D systems.

**3.1.2.1. 2D quadruple topolectrical circuits and higher-order Anderson topolectrical circuits**

Quantized multipole insulators, first proposed in electronic systems, introduces the idea of topologically protected corner states that carry fractional charge [87, 88]. This concept was extended to circuit lattices by Imhof et al., who have realized a quadrupole topological insulator in a 2D topolectrical circuit [89]. They present a schematic diagram of this 2D quadrupole circuit, where the circuit nodes are arranged in a square lattice with specific coupling configurations that generate the quadrupole moment. The circuit equations yield an admittance matrix:

$$\tilde{J}(\omega_0, \vec{k}) = \sum_n e^{-i\vec{k}\cdot a_n} J_{0a_n}(\omega_0)$$
$$= i\sqrt{\frac{c}{l}}\left[(1+\lambda\cos k_x)\sigma_1\tau_0\right] + i\sqrt{\frac{c}{l}}(1+\lambda\cos k_y)\sigma_2\tau_2 - i\sqrt{\frac{c}{l}}\lambda\sin k_x\sigma_2\tau_3 + i\sqrt{\frac{c}{l}}\lambda\sin k_y\sigma_2\tau_1.$$

(36)

This matrix form aligns with the Hamiltonian of a quadruple topological insulator, thereby allowing the circuit to simulate the corresponding topological properties. The impedance measurements confirmed the presence of zero-dimensional corner states in the 2D circuit network, validating the simulation of the quadruple topological insulator's characteristics. Moreover, the quadrupole transitions and edge mode topology have also been observed in an LC circuit network designed by Serra-Garcia et al [90]. Except for the quadruple topological insulator, Wu et al have designed a kagome circuit with capacitors and inductors, and observed the corner states that exist in the bulk gap, the continuum of bulk modes, and the dependence on the corner shapes by measuring the resonance peaks of the corner impedance [91].

On the other hand, it is generally believed that strong disorder can close the energy gap in topological systems, thus destroying the topological phase. Contrary to this conventional understanding, it has been theoretically demonstrated that disorder could induce the formation of topological states, which they named Anderson topological insulators [92]. The theoretical demonstration on the realization of Anderson topological insulators has been proposed by Zhang et al [93]. With the advancement in research on higher-order topological insulators and Anderson topological insulators, an important question has emerged: Do higher-order Anderson topological insulators exist, and if so, how can they be experimentally verified? Zhang et al., utilizing a modified Haldane model with additional lattice distortions, demonstrated the existence of higher-order Anderson topological phases by introducing disorder in the next-nearest-neighbor coupling phases [24], as presented in Figures 6a and 6b. Then, researchers combined 'woven' and 'non-woven' circuit network node couplings to effectively implement real-valued next-nearest-neighbor and complex-valued nearest-neighbor lattice couplings. The schematic of the designed sample is shown in Figure 6c. Furthermore, by controlling the ratio of disordered 'woven' connections in the circuit network nodes, researchers introduced different strengths of next-nearest-neighbor coupling disorder into the designed circuit system. As the disorder strength increased, zero-dimensional corner states induced by disorder were observed, as shown in Figure 6d, confirming the existence of higher-order Anderson topological phases.

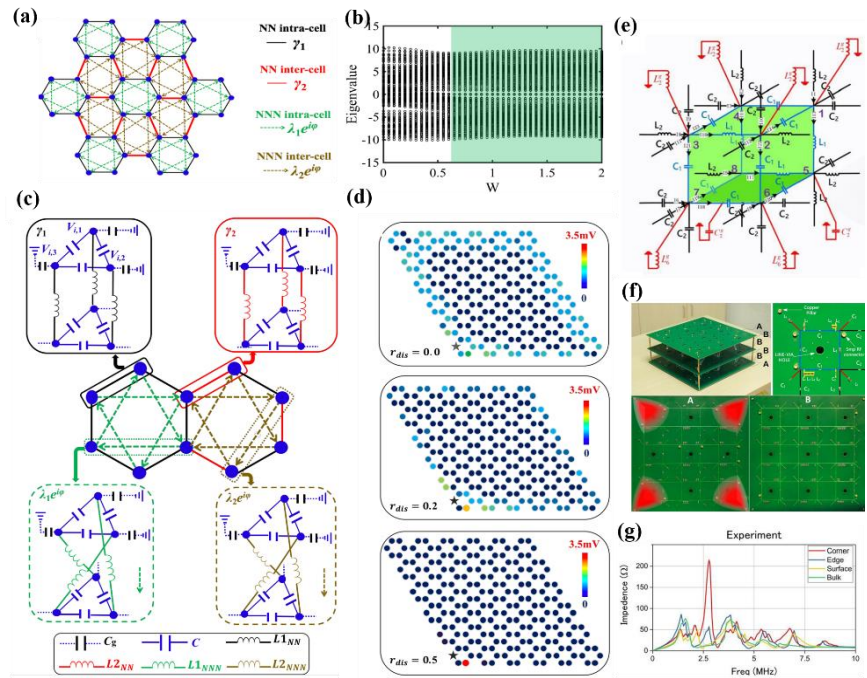

**Figure 6. Second-order Anderson topolectrical circuits and higher-order topolectrical circuits.** (a). The scheme of the extended Haldane model. (b). Eigen-spectra of the extended Haldane model with different strengths of disorder for the next-neighbor coupling phases. (c). The schematic diagram of higher-order Anderson topological circuit. (d). Measured voltage distributions with different strengths of disorder for the next-neighbor coupling phases. Figures a-d are taken from [24]. Reproduced with permission [24] Copyright 2021, American Physical Society. (e) and (f). The schematic diagram and fabricated sample of 3D octupole topolectrical circuits. (g). Measured impedance spectra for the 3D octupole circuit. Figures e-g are taken from [94]. Reproduced with permission [94] Copyright 2019, American Physical Society.

### 3.1.2.2. 3D Octupole and 4D Hexadecapole topolectrical circuits

Since the theoretical proposal for higher-order topological states was introduced in 2017, reports of observing such states in quantum materials and classical artificial systems have gradually emerged. However, due to numerous experimental constraints, initial studies were confined to exploring higher-order topological phases in two-dimensional systems. Realizing higher-order topological states in three dimensions or higher poses significant challenges. It is noted that the properties of electric circuits are determined by the way circuit network nodes are connected, independent of the circuit's shape or spatial dimensions. This makes it feasible to realize quantum states that are difficult to achieve in condensed matter and other classical systems within circuit systems.

By constructing anticommuting mirror symmetries along the $x$, $y$, and $z$ directions, a three-dimensional octupole topological phase induced by bulk octupoles can be generated. This phase supports topologically protected zero-dimensional corner states. The classical analogue of the three-dimensional octupole topological phase requires constructing both positive and negative lattice couplings in real space, a stringent requirement that can be met through capacitive and inductive connections between circuit grid nodes. Bao et al. have demonstrated the realization of a three-dimensional octupole topological phase using classical circuits [94], with the circuit schematic shown in Figure 6e. The fabricated circuit sample and its local magnified view are shown in Figure 6f. Impedance responses measured at the circuit's corner nodes confirm the presence of higher-order topological corner states within the energy gap, as presented in Figure 6g. Liu et al have designed a similar 3D topological circuit for realizing 0D corner states of octupole topological insulators [95].

Higher-dimensional topological states often exhibit a range of novel physical properties that do not exist in low-dimensional systems. However, due to the limitations of three-dimensional space, both naturally occurring quantum materials and artificially designed classical structures struggle to realize topological states beyond three dimensions. The properties of classical circuits are determined by the connectivity of the circuit network nodes, independent of the specific shape or spatial dimensions of the network. Therefore, higher-dimensional circuit networks can be projected onto a two-dimensional plane. Building on this advantage, Zhang et al have designed a classical circuit analogue for a 4D hexadecapole topological insulator. By comparing numerical simulations and experimental measurements of topological and trivial circuits, zero-dimensional corner states induced by the four-dimensional hexadecapole were verified.

### 3.1.2.3. Other types of higher-order topolectrical circuits

Various other higher-order topological states have been realized using topolectrical circuits. These include non-Abelian higher-order topological bound states in the continuum [96], Z-class higher-order topolectrical circuits [97], quasicrystal HOTIs [98, 99], HOTIs generalized from SSH models [100], higher-order topolectrical circuits implemented in kagome and hybrid honeycomb-kagome lattices [101, 102], higher-order TSMs [29], square-root higher-order topological insulator in electric circuits [1012], HOTIs in bilayer asymmetric SSH topological electric circuits [103], higher-order topolectrical circuits with type-II and type-II corner states [104, 105] and others [106-109]. Each of these realizations demonstrates the versatility of topolectrical circuits as a platform for exploring and understanding higher-order topological phases.

### 3.2. Engineering non-Hermitian topological states in electric circuit networks

While the realization of Hermitian topological phases in circuits has significantly advanced our understanding of topological states, the exploration of non-Hermitian topological circuits introduces new complexity and potential. Non-Hermitian systems, characterized by their non-conservative nature and the presence of gain and loss, exhibit unique phenomena that do not appear in Hermitian systems. In this section, we summarize how non-Hermitian topological circuits can be designed and implemented, highlighting their distinctive properties and experimental realizations.

**3.2.1. 1D non-Hermitian SSH topolectrical circuits and topological circuit sensors**

In lattice models, non-reciprocal hopping terms can be realized in circuits using a negative impedance converter through current inversion (INIC). This setup allows for asymmetric hopping, where hopping in one direction is negative, while in the opposite direction, it is positive (as discussed in Section 2). Building upon the non-reciprocal couplings facilitated by INICs, a 1D non-Hermitian SSH model can be conveniently implemented within circuit systems. For example, Helbig et al. have demonstrated a non-Hermitian SSH circuit model using non-reciprocal intra-cell couplings enabled by INICs [25]. The circuit and its associated equations align well with the theoretical framework of the non-Hermitian lattice model. Through the use of the admittance matrix method, the non-Hermitian eigen-spectra with open and periodic boundaries were observed in the circuit. Similar results are also reported in Ref. [110, 111], where the non-Hermitian bulk-boundary correspondence explained by the non-Bloch winding number has been explored in electric circuits. In addition, based on the interplay between topological states and non-Hermitian skin effect, Lu et al have realized the extended midgap topological states in the non-Hermitian 1D SSH electric circuits [112].

Moreover, recent theoretical advances reveal that non-Hermitian topological boundary states exhibit an exponential shift with increasing system size when subjected to external perturbations [113]. Yuan et al. have harnessed this phenomenon to design a non-Hermitian topological circuit sensor characterized by high sensitivity and robustness [114], as depicted in Figure 7a. The eigenfrequency spectrum of the circuit, shown in Figure 7b, confirmed the presence of topological zero-energy modes, akin to those in the non-Hermitian SSH model. To validate the circuit's performance, the researchers introduced boundary perturbations and analyzed the frequency shifts of the topological zero-energy modes under varying levels of disorder (Figure 7c). The results demonstrate that with increasing circuit size, the frequency shift due to boundary perturbations exhibits an exponential growth, with this sensitivity proving highly resistant to disorder. Additionally, four different lengths of non-Hermitian topological circuits were fabricated. By measuring the impedance response, it was shown that the frequency shift of the impedance peak, associated with the topological zero-energy mode, significantly increases with circuit length. The researchers further integrated displacement, rotation angle, and liquid-level-dependent capacitors into the circuit. Connecting these capacitors to the circuit's ends enabled ultra-sensitive detection of minute variations in physical quantities.

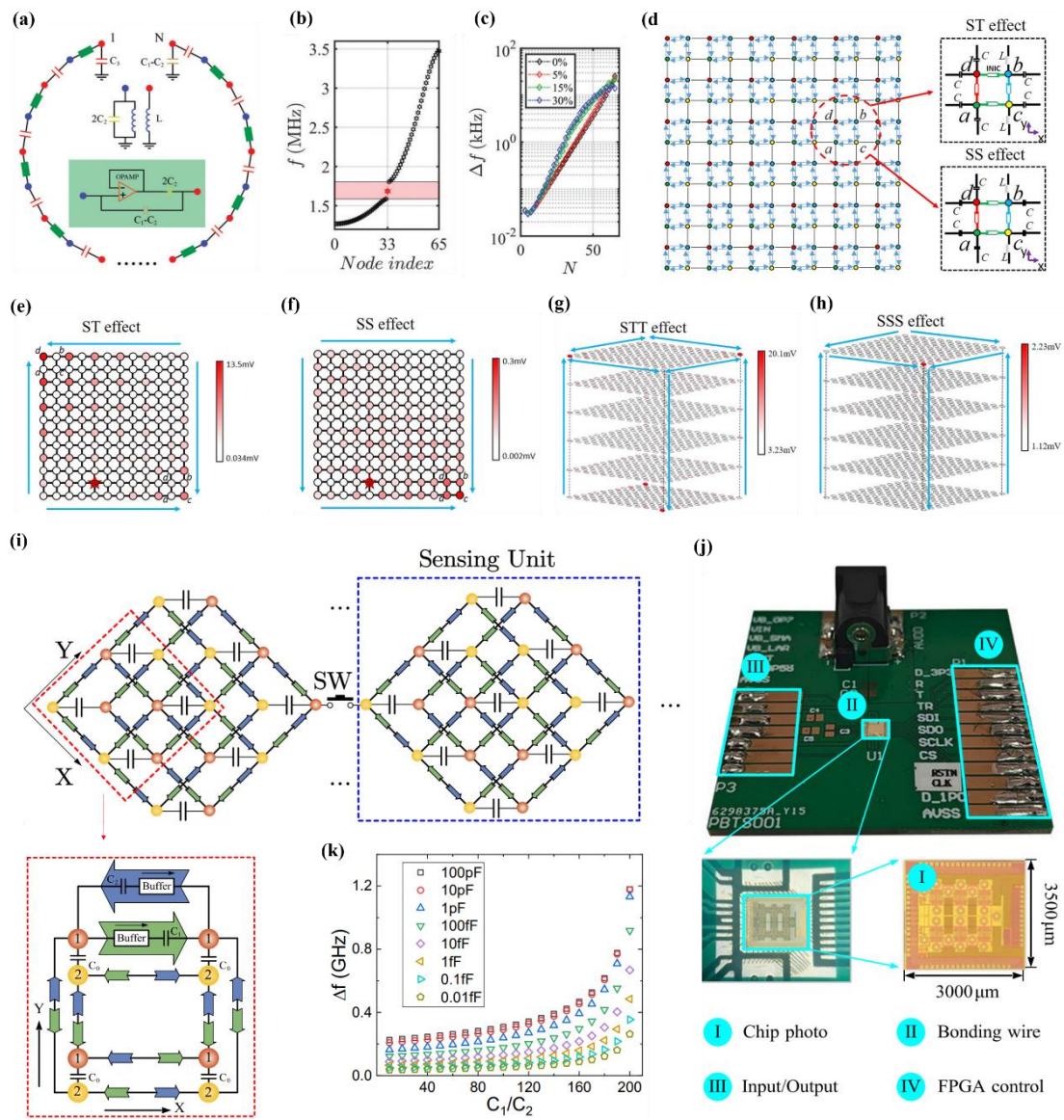

**Figure 7. Non-Hermitian topolectrical circuits.** (a). The scheme of 1D non-Hermitian topological circuit sensors. (b). The calculated eigen-spectrum of 1D non-Hermitian topological circuit with open boundary condition. (c). The frequency shifts of the topological zero-energy modes with respect to boundary perturbations under varying levels of disorder. Figures a-c are taken from [114]. Figures are adapted from CC-BY open access publications Wiley-VCH Verlag. (d). The schematic diagram of 2D non-Hermitian circuit model, comprising LC circuits and INICs. (e) and (f). Measured static voltage distributions for 2D non-Hermitian circuits with skin-topological and skin-skin effects. (g) and (h). Measured static voltage distributions for 3D non-Hermitian circuits with skin-topological-topological and skin-skin-skin effects. Figures d-h are taken from [116]. Figures are adapted from CC-BY open access publications Springer Nature. (i). 2D integrated-circuit sensor. (j). The photon of the fully integrated circuit chip fabricated using a standard 65nm complementary metal oxide semiconductor process technology. Figures i and j are taken from [120]. Figures are adapted from CC-BY open access publications AAAS.

### 3.2.2. 2D and 3D non-Hermitian SSH topolectrical circuits with hybridized skin-topological modes

The interplay between topological boundary states and the non-Hermitian skin effect has given rise to novel hybrid states [115]. Zou et al. have provided the first experimental realization of hybrid higher-order skin-topological states in 2D and 3D topolectrical circuit systems [116]. The 2D circuit model, comprising LC circuits and INICs, is illustrated in Figure 7d. By carefully engineering the inter-site couplings within the circuit Laplacian, both skin-topological and skin-skin effects were observed in 2D non-Hermitian SSH topolectrical circuits by measuring static voltage distributions, as shown in Figures 7e and 7f. Additionally, as presented in Figures 7g and 7h, the study extended these observations to 3D non-Hermitian SSH topolectrical circuits, where the interaction between skin modes and topological effects was similarly observed, showcasing the versatility and richness of these hybrid states in higher-dimensional systems [117-119].

Moreover, Deng et al. have theoretically designed and experimentally demonstrated a 2D

integrated-circuit sensor with exceptional performance based on the exotic properties of high-order non-Hermitian topological physics [120], as depicted in Figure 7i. The frequency shift induced by perturbations in these sensors exhibits exponential growth relative to the device size, surpassing the limitations of conventional sensors. Figure 7j presents the photon of the fully integrated circuit chip fabricated using a standard 65nm complementary metal oxide semiconductor process technology. Experimental verification has confirmed that these systems not only exhibit sensitivity below 1*fF*, as illustrated in Figure 7k, but also demonstrate robustness against disorders. This ultra-sensitive integrated circuit sensor holds immense potential for diverse applications across various fields and presents an exciting prospect for next-generation sensing technologies.

### 3.2.3. Other non-Hermitian topolectrical circuits

In addition to the aforementioned studies, several significant advancements have been made in the field of non-Hermitian topological circuits. Wu et al. have observed non-Bloch dynamics through a temporal topolectrical circuit, providing new insights into the dynamic behavior of these systems [121]. The simulation of domain-wall non-Hermitian topological interface states has been achieved using both 1D electric circuits [122] and high-dimensional circuit networks [123], demonstrating the versatility of topolectrical circuits in modeling complex topological phenomena. Further, Hofmann et al. have reported the discovery of a reciprocal skin effect in circuit lattices [124]. Cao et al have observed the different knot topologies around EPs by tunable non-Hermitian topological circuits [125]. Other properties of exceptional points have also been widely explored in electric circuits [118, 126-129]. Wu et al have reported the circuit realization of loss-induced non-Hermitian second-order topological phases [130]. Zou et. al reported the realization of exceptional bound states by electric circuit [117, 131-133]. Stegmaier. et al observed Topological Defect modes and PT-symmetry in Non-Hermitian Electrical Circuits [134]. The circuit design for the realization impurity induced scale-free localization has been proposed by Li et. al [135]. Zheng et al have realized several topological properties of 5D non-Hermitian systems by circuit networks, including Yang mono-spheres, Fermi cylinder surfaces and Fermi arcs connecting the two Yang mono-spheres [136].

Zhang et al. have reported a topological switch capable of controlling the on-off skin effect, introducing a new level of control in non-Hermitian circuits [137]. The exploration of mirror-symmetric skin effects is carried out by Yoshida et al., who provided valuable insights into how symmetry influences the non-Hermitian skin effect [138]. Critical hybridization of skin modes in coupled non-Hermitian circuit has also been proposed [139]. Zhu et al. have demonstrated the observation of higher rank chirality and non-Hermitian skin effect in a topolectrical circuit [140]. Zhang et al. have reported an experimental observation of continuum Landau modes in non-Hermitian electric circuits, in which the non-Hermitian Dirac Hamiltonian is simulated by non-reciprocal hoppings and the pseudomagnetic field is introduced by inhomogeneous complex on-site potentials [141]. Moreover, Zhu et al. have reported the experimental observation of the rank-2 skin effect, characterized by the coexistence of edge and corner localized skin modes, marking a significant step forward in understanding higher-order topological phenomena in non-Hermitian systems. Lee et al have studied many types of skin effect which rely on different couplings and symmetries [142-151]. These studies collectively underscore the rapidly evolving landscape of non-Hermitian topological circuits, opening new avenues for research and application.

### 3.3. Nonlinear and time-varying topolectrical circuits

In addition to linear topolectrical circuits, nonlinear topolectrical circuits have garnered significant attention due to their unique properties and potential applications. Hadad et al. have demonstrated the circuit implementation of the 1D nonlinear SSH model using voltage-dependent varactors [152]. It is shown that nonlinear circuit arrays can exhibit self-induced topological transitions as a function of the input intensity, leading to topologically robust edge states that are immune to the presence of defects. Wang et al. have advanced the field by utilizing a nonlinear microwave topological circuit (Figure 8a) to achieve topologically protected harmonic generation [153]. Kotwal et al. have taken a different approach by integrating nonlinear Chua's diodes with topological circuits, leading to the experimental realization of nonlinear topological self-oscillations, which offered new insights into the interplay between nonlinearity and topology [154]. Based on the interplay between the 1D SSH topology and nonlinearity [27], Hohmann et al have observed boundary-localized cnoidal waves in a nonlinear topological circuit (Figure 8b), where the circuit equation follows the Korteweg-de Vries equation. Ventra et al. have further expanded the scope of this research by being the first to observe topological states protected by supervised chiral symmetry in a memristor-based topological circuit, marking a significant milestone in the exploration of nonlinear topoelectrical phases [155]. Moreover, novel properties of nonlinear exceptional points have also been explored in electric circuits [156-161].

On the other hand, time-varying topoelectrical circuits have emerged as a compelling area of study, particularly for their ability to leverage dynamic processes to engineer new topological states. A prime example is the work on topological pumping circuits [162], as plotted in Figure 8c, where the adiabatic modulation of system parameters is shown to induce quantized center-of-mass transport. Further advancing this domain, a recent study has delved into the nucleation of topological edge states in frequency space, realized through Floquet electrical circuits [163]. Additionally, the synergistic engineering of both temporal and spatial degrees in circuit networks holds tremendous potential across diverse technologies. Zhang et al have proposed topoelectrical space-time circuits (Figure 8d) to fulfill this functionality. By designing and applying a novel time-varying circuit element controlled by external voltages, three distinct types of topological space-time crystals are experimentally demonstrated, including the (1+1)-dimensional topological space-time crystal with midgap edge modes, (2+1)-dimensional topological space-time crystal with chiral edge states, and (3+1)-dimensional Weyl space-time semimetals. Moreover, Floquet topological states have been realized in an integrated circuit platform [164], showing potential applications in wireless Communication. These investigations highlight the potential of periodically driven circuits to create and manipulate novel topological phenomena, significantly enriching the field. Finally, in addition to time-varying circuits, static topological circuits have also been designed to directly explore the Floquet topological states in the synthetic frequency space [165, 166].

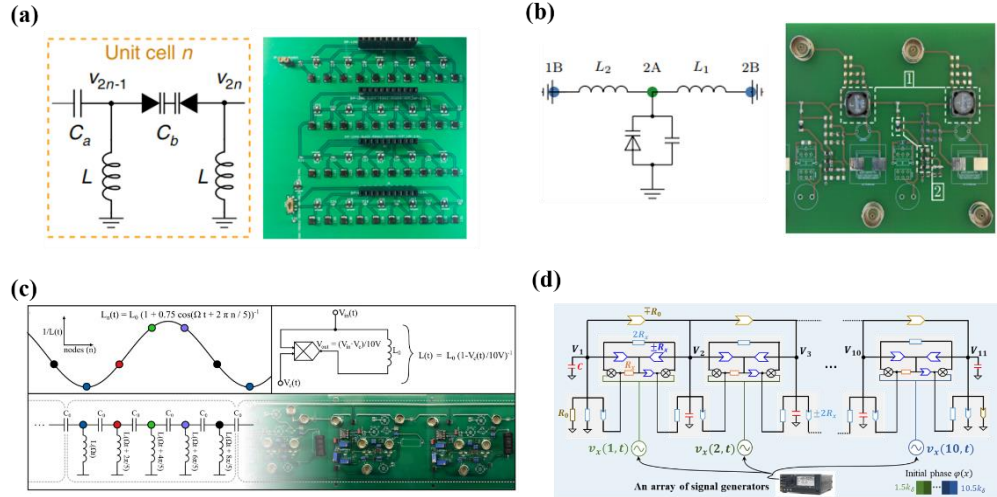

**Figure 8. Nonlinear and time-varying topolectrical circuits.** (a). A nonlinear microwave topological circuit with topologically protected harmonic generation. Figure a is taken from [153]. Figure is adapted from CC-BY open access publications Springer Nature. (b). Active nonlinear topological circuit with Chua's diodes. Figure b is taken from [27]. Figure is adapted from CC-BY open access publications American Physical Society. (c). 1D topological pumping circuits. Figure b is taken from [162]. Figure is adapted from CC-BY open access publications American Physical Society. (d). Space-time topological circuits.

## 3.4. Simulating few-body quantum systems by circuit networks

By mapping the Schrödinger equation of low-dimensional few-body quantum systems onto the eigen-equations of electrical circuits embedded in high-dimensional networks, it becomes possible to simulate the complex dynamics of strongly interacting few-body quantum systems using classical electrical circuits. This innovative approach leverages the versatility and accessibility of topolectrical circuits, allowing for the detailed study of quantum phenomena that are otherwise challenging to investigate directly.

Ren et al. have studied the Berry phase of the interacting open system based on the electric circuit [167]. Olekhno et al. have designed and fabricated a two-dimensional circuit (Figure 9a) to simulate strong interaction-induced two-body topological states, demonstrating the capability of topolectrical circuits to model complex quantum behaviors [168]. Few-body interactions with higher-dimensional circuits are also presented [169-172]. Building on this method, Zhang et al. have experimentally simulated Dirac-like phenomena in two strongly correlated bosons by 2D circuits, including Zitterbewegung and Klein tunneling [173]. In this study, the eigenstates of two correlated bosons were mapped onto modes of a specially designed circuit lattice (Figure 9b), with interaction-induced Zitterbewegung and Klein tunneling verified through voltage dynamics measurements. Moreover, Zhou et al. have extended this approach by directly mapping the eigenstates of two correlated bosons in one-dimensional Aharonov-Bohm cages onto modes of a two-dimensional circuit lattice. This allowed them to experimentally simulate interaction-induced flat-band localizations and topological edge states in 2D electrical circuit networks (Figure 9c) [174]. By adjusting the effective interaction strength through circuit grounding, they detected two-boson flat bands and topological edge states by measuring frequency-dependent impedance responses and voltage dynamics over time. In addition to the two-boson systems, this method has been extended to few-boson, and the interaction induced boundary-localized bound states in the continuum has been simulated by circuit networks [175].

Except for interacting bosonic systems, electric circuits are further engineered to simulate two-anyon systems. Zhang et al. have reported the experimental simulation of anyonic Bloch oscillations using two-dimensional electric circuits (Figure 9d) [176]. By mapping the eigenstates of two anyons onto the modes of designed circuit simulators, they confirmed the Bloch oscillations of two bosons and two pseudofermions through voltage dynamics measurements. Notably, the oscillation period in the two-boson simulator was nearly twice that in the two-pseudofermion simulator, consistent with theoretical predictions. Building on this work, Zhang et al. have further revealed and experimentally simulated a type of bound state in the continuum in anyonic systems using electric circuits (Figure 9e), where the eigenstates of two anyons were mapped onto modes of the designed circuit networks [177]. Additionally, Olekhno et al. have reported the simulation of topological transitions driven by quantum statistics using electrical circuits [178]. The exotic Bloch oscillations (BOs) induced by the non-Abelian fusion of Fibonacci anyons were also successfully simulated in electric circuits by Zhou et al [179].

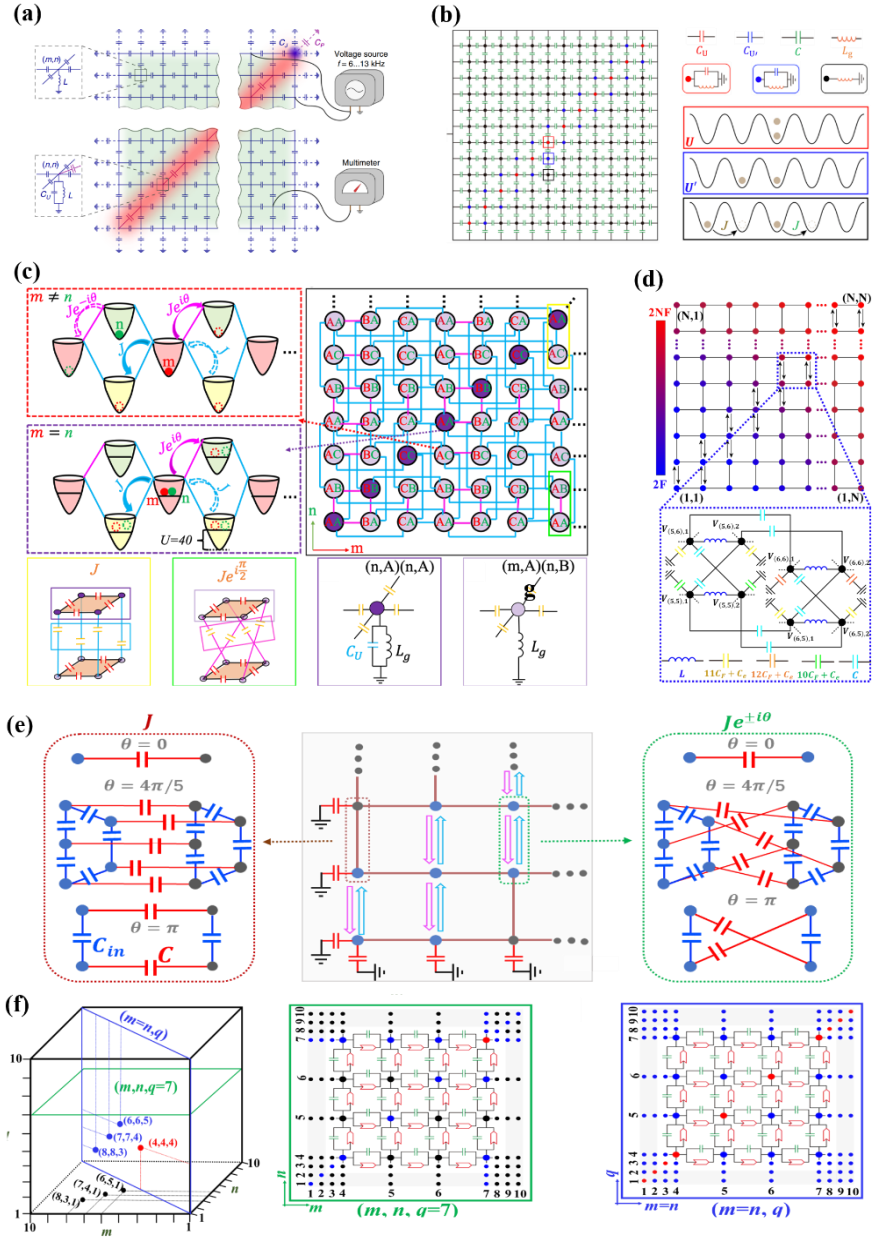

**Figure 9. Simulating few-body quantum systems by circuit networks** (a), (b) and (c). 2D electric circuits for simulating interaction two-body topological edge states, interaction-induced two-body Dirac physics and interaction-induced two-boson AB cage, respectively. Figure a is taken from [168]. Figure is adapted from CC-BY open access publications Springer Nature. Figure b is taken from [173]. Figure is adapted from CC-BY open access publications Springer Nature. Figure c is taken from [174]. Reproduced with permission [174] Copyright 2023, American Physical Society. (d), (e) and (f). The 3D non-Hermitian circuits for simulating interaction-induced non-Hermitian aggregation effects of three bosons. Figure d is taken from [176]. Figure is adapted from CC-BY open access publications Springer Nature. Figure e is taken from [177]. Figure is adapted from CC-BY open access publications Springer Nature. Figure f is taken from [180]. Reproduced with permission [180] Copyright 2022, American Physical Society.

Finally, the non-Hermitian skin effects induced by strong interactions were simulated using non-Hermitian electric circuits (Figure 9f) [180]. This interaction-induced non-Hermitian effect, known as the non-Hermitian aggregation effect, represents the clustering of bosonic states with differing occupations in a periodic lattice. By mapping the eigenstates of three correlated bosons onto modes of a three-dimensional electric circuit, the interaction-induced non-Hermitian aggregation effects in Hilbert space were verified by measuring the spatial impedance response. The significance of these advancements lies in their ability to extend our understanding of quantum systems by using accessible, classical tools to explore complex quantum phenomena. The versatility of topolectrical circuits enables researchers to model and experimentally verify a wide range of quantum phenomena states, including those influenced by strong interactions, anyonic statistics, and non-Hermitian effects. These insights not only deepen our fundamental knowledge but also open up new avenues for technological applications in quantum information processing, materials science, and beyond.

## 3.5 Experimental observation of classical analogy of topological entanglement entropy

The topological order is responsible for many fundamental phenomena [181-186], e.g., the fractional quantum Hall effect, spin liquids and, etc. As an exotic quantum phase, topological order cannot be understood by conventionally local order parameters. Instead, the topological order represents an area law topological phase while maintaining the long-range entanglement. Nontrivial properties can be uncovered in topological orders, such as topological degeneracy, fractional statistics, topological entanglement entropy (TEE), and so on. Wen, et al. have pointed out that on the high-genus Riemann surface, the ground state degeneracy is directly related to the statistics of the quasi particles [181]. Arovas, et al. have shown that the excitations of quasi particles entering the quantum Hall effect obey the fractional statics, and connect to the fractional charge directly [182]. In addition, Wen has revealed that the edge excitations in the chiral Luttinger liquids reflect the topological order in the bulk [183]. For the two-dimensional topologically ordered medium, the many-particle quantum entanglement in the ground state can be characterized in a universal way. Kitaev, Preskill, Levin and Wen point out this characterized value is just the TEE, which is closely related to the superselection sectors of the medium [184, 185]. A prominent example to illustrate how to obtain the TEE is the toric code model, which is proposed by Kitaev [186]. The toric code model can exhibit the $Z_2$ topological order, whose Hamiltonian is shown as $H_0 = -\sum_s A_s - \sum_p B_p$, where the elements $A_s = \prod_{i \in s} \sigma_i^x$ and $B_p = \prod_{i \in p} \sigma_i^z$, in which the subscripts $s$ ($p$) denote the vertices (plaquettes) in a square lattice. An efficient method to characterize the topological order is using the entanglement of the ground state $|\varphi_g\rangle_i$ ($i = 1, 2, \cdots n$), which is just the TEE. The value of TEE is obtained from the entropy of system with different scales. By obtaining the entropy of system, it is commonly to separate the system into A and B two subsystems. The relation between the entropy and the length $L_x$ is presented as $S(A) = \alpha L_x - m\gamma + \ldots$. In the expression, the quantity $\gamma$ is just the TEE. The TEE for the toric code with $Z_2$ topological order is $\gamma = \ln 2 = 0.6932$. Not only for the toric code model, other many-body systems can show non-trivial topological order and non-zero TEE values. A practical method to identify topological orders in arbitrary realistic models is using the numerical calculation. The density matrix renormalization group (DMRG) is a powerful numerical technique, which can accurately calculate the TEE for many different models. Based on this DMRG method, Jiang, et al. [187] have studied many different kinds of Hamiltonians including toric code model in the magnetic fields and the quantum S=1/2 antiferromagnet on the Kagome lattice.

Not only in the theoretic and numerical studies, the experimental demonstration of this topological order also attracts lots of attentions. Han, et al. have provided a possible experimental proposal based on quantum circuits [188]. Following this experimental proposal, by creating the ground state and excited states of the Kitaev Hamiltonian, Lu, et al. have implemented the anyonic braiding and fusion operations within six-qubit graph states by single-qubit rotations [189]. In addition, by encoding the many-body state of the toric code to the multi-partite entangled polarized photons, Pachos, et al. have revealed the characteristic fractional statistics of anionic states [190]. The experiment demonstration of the behaviors in the topological order system is not limited to the optical platform. Based on the nuclear spins, Kitaev, et al. have demonstrated three topologically ordered matter phases including toric code, doubled semion, and doubled Fibonacci model, by measuring their modular *S* and *T* matrices [191-194]. Even though the presence of generic disorders and detuning from the exactly solvable point of the toric code, Luo, et al. have reconstructed the modular matrices in the good accuracy by experimental techniques to uncover the topological order [195]. Not only for the toric code, Peng, et al. have realized the experimental quantum simulation of the Wen-plaquette spin model on the nuclear spins [196]. In addition, with the superconducting circuit platform and the ultracold atom platform, the toric code Hamiltonians containing four-spin plaquette arrays are realized, and the 1/2 mutual statistics during the braiding process are also demonstrated [197, 198].

However, the experimental demonstrations above only focus on the braiding statistics, and do not consider the TEE. The key obstacle to observe the TEE is the difficulty in constructing the ground state of the large-scale toric code model. Due to the difficulty in the direct construction of the large quantum system to observe the TEE, the corresponding experimental demonstrations are barely.

Recently, Chen et al. have used the classical analog scheme to demonstrate the TEE [199]. With the aid MESs, the topological order of the toric code is characterized exactly in experiment. The nontrivial topological order embodied in the ground states of the toric code is experimentally demonstrated, and the phase transition from $Z_2$ order phase to trivial phase in the toric code is also simulated. The key to realize such the observations is the good mapping from the classical signals in the circuit to the correlation measurement in the quantum experiments. In this microwave circuit experiment, the classical signals simulate the MESs exactly. The details of experimental setup are shown in Figure 10a and 10b.

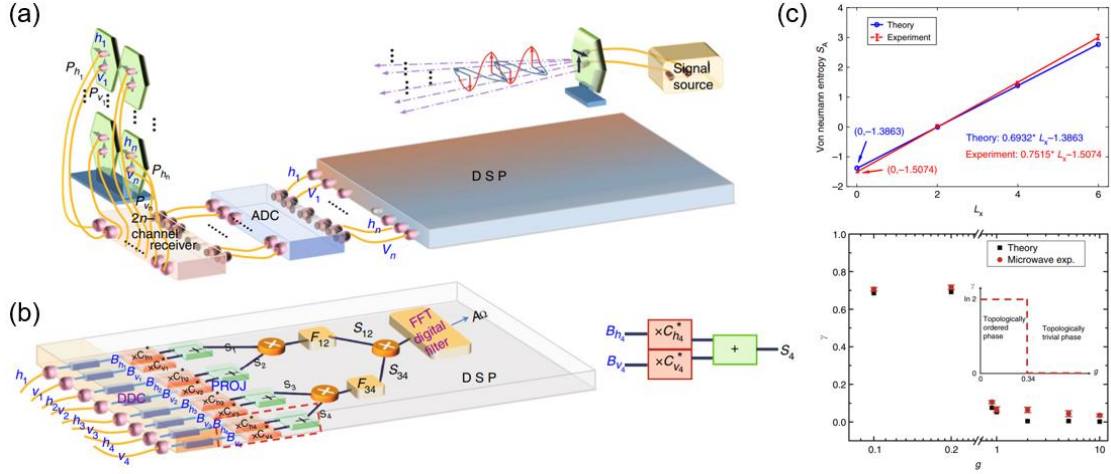

**Figure 10**. The experimental setup to construct the CMESs as the analogy of MESs. (a) The designed classical microwave signal processing system. (b) The construction of classical analog of the 4-qubit MES. (c) The experimental demonstration of topological order and transition in the toric code model. These figures are taken from [199]. Figures are adapted from CC-BY open access publications Springer Nature.

Based on the same experimental setup, the experimental results for the "classical minimal entropy states" (CMESs) with four, eight and twelve spins are also obtained. The corresponding entropies of the CMESs are addressed as the red dotted lines in Figure 10c. With these entropy values in experiment, the TEE is extracted as the value $\gamma = 0.7537$, and this value is very close to that shown in quantum theory. Within the results of TEE in experiment, it has already demonstrated the successful observation of the classical analog of topological order. The topological transition from the $Z_2$ order to the trivial phase can also reflected in the same way. It no doubt demonstrates that the effectiveness of this classical platform to study the topological order.

Following the study of entanglement entropy to characterize topological order in many-body system, Maiellaro, et al. has extended to the edge quantum conditional mutual information and edge squashed entanglement, which can be used as the valid detectors for topological superconductors, like the Kitaev spin chain with long-range hoppings, featuring geometrical frustration and a suppressed bulk-edge correspondence [200]. And for fractals and quasicrystals, Manna, et al. have used the quasiparticle properties to detect quantum phase transitions [201].

The idea to realize classical analog of topologic order has also been applied to observe the salient signatures of topological phases in entanglement entropy and entanglement spectrum, which has been proposed by Lin, et al. [202]. With the obtained nonlocal correlations, the Gioev-Klich-Widom scaling law in phononic systems are verified.

## 4. Topological Quantum Computing with Classical Circuits

The exploration of topological states holds significant importance in the pursuit of achieving topological quantum computing. Quantum computers manufactured by fully utilizing the unique quantum properties have significant advantages over classical computers in solving some problems [1-5]. But quantum states are highly susceptible to environmental interference, leading to decoherence effects and errors of quantum gates caused by limited quantum operation accuracy. Until now, there has been no implementation of a practical universal quantum computer. At present, there are two propositions to solve these problems. One way is to use logical qubits and error-correcting codes to correct all errors that appear during computation [6-14]. Such a scheme is extremely difficult to implement because it requires a lot of additional resources for quantum circuits. Another way is to perform a topological quantum computing scheme.

Topological quantum computation aims to employ anyonic quasiparticles with exotic braiding statistics to encode and manipulate quantum information in a fault-tolerant way [186, 203-206]. Majorana zero modes are experimentally the simplest realization of anyons that can non-trivially process quantum information. Based on the scheme of Majorana zero modes, some potential realizable systems have been analyzed theoretically, such as superconductor and fractional quantum Hall liquid [207-214], semiconducting heterostructures and wires [215-220], photonic networks [221-223] and so on. However, the experimental implementation has encountered great challenges, although some simulations of braiding zero modes can be accomplished in acoustics and photonic systems experimentally [224, 225]. The simulation of topological quantum computation requires to braid zero modes many times, which is beyond the capability of these physically classical simulators.

Fortunately, due to the mature technology and direct manipulation of classical circuits, it is expected that the Majorana zero mode can be created and deeply manipulated. In 2020, Ezawa theoretically proposes the use of classical circuits to demonstrate the Majorana-like zero mode and braiding process to form quantum gates [34]. But, in this scheme, not only do the variable capacitor and inductor need to be precisely regulated at the same time, but also the whole braiding operation needs to be completed in a very short time. This is because the input signal decays quickly due to the loss of circuit components. Therefore, it is very difficult to accomplish in experiments. In 2023, Zou et al. have proposed a new classical circuit scheme to overcome the problems of the previous schemes [226]. Based on such classical circuits, the generation of Majorana-like zero modes, quantum gates formed by braiding process, and functions of topological quantum computing were experimentally simulated. Below, the circuit simulation of topological quantum computing is introduced.

**4.1. Theorical design of braiding process in classical circuits**

The key steps to achieve topological quantum computation is to construct Majorana zero modes and braiding process. The construction and manipulation of Majorana zero modes can be achieved by designing the Kitaev model in classical circuits, as discussed in Section 3.1.1.1. Subsequently, we present an introduction to the braiding process in classical circuits.

The purpose of the braiding process is to facilitate the exchange of a pair of Majorana edge states, which can be accomplished by utilizing a T junction based on the Kitaev chain. In Ref. [34], Ezawa theoretically proposed the use of classical circuits to demonstrate the T junction for braiding process. The circuit diagram consists of two parts: one part is a finite Kitaev chain which can achieve the movement of Majorana edge states by controlling the ground connection; the other part is a finite Kitaev chain with phase $\phi$ which can achieve both the movement (by controlling the ground connection) and phase rotation (by controlling the phase $\phi$) of Majorana edge states. The corresponding circuit Laplacian $J_T$ for the T junction can be expressed as:

$$J_T = \begin{pmatrix} h_1 & g_1 \\ g_2 & h_2 \end{pmatrix} \tag{37}$$

with

$$h_1 = -2C\cos k + 2C - \left(\omega^2 L_0\right)^{-1},$$

$$h_2 = 2\left(\omega^2 L\right)^{-1} \cos k - 2\left(\omega^2 L\right)^{-1} + C_0,$$

$$g_1 = -C_X e^{ik} + \left(\omega^2 L_X\right)^{-1} e^{-ik},$$

$$g_2 = -C_X e^{-ik} + \left(\omega^2 L_X\right)^{-1} e^{ik}, \tag{38}$$

where $k$ is the wavevector and $\omega$ is the angular frequency. This circuit Laplacian closely mirrors the Hamiltonian of the T junction. The entire braiding process, based on the aforementioned T junction, comprises eight steps that involve rotating phases and moving two edge states. The two topological edges undergo an exchange after the braiding process.

## 4.2. Experimental simulation of topological quantum computing in classical circuits

Although the Majorana zero modes and braiding process is theorical designed in circuit system above, it is very difficult to accomplish in experiments. There are two main reasons: a) the requirement that the variable capacitor and inductor need to be precisely regulated at the same time. b) the requirement that the whole braiding process needs to be completed in a short time due to the quickly decrement of the input signal. How to experimentally simulate topological quantum computation still needs to be explored.

Recently, in Ref. [226], not only theorical design, but also experimentally simulation of braiding process, gate operation and Grover's algorithm have been achieved in circuit system. In this work, a new classical circuit scheme has been designed to overcome the problems of the above scheme. Resistor–capacitor circuit instead of inductor–capacitor circuit has been constructed. It has the advantage that only the resistance needs to be adjusted during the braiding process, without the need to precisely adjust the capacitance and inductance at the same time. In addition, segmented fixed resistances instead of variable resistances are used, which allows the whole braiding process to be completed before the signal is attenuated.

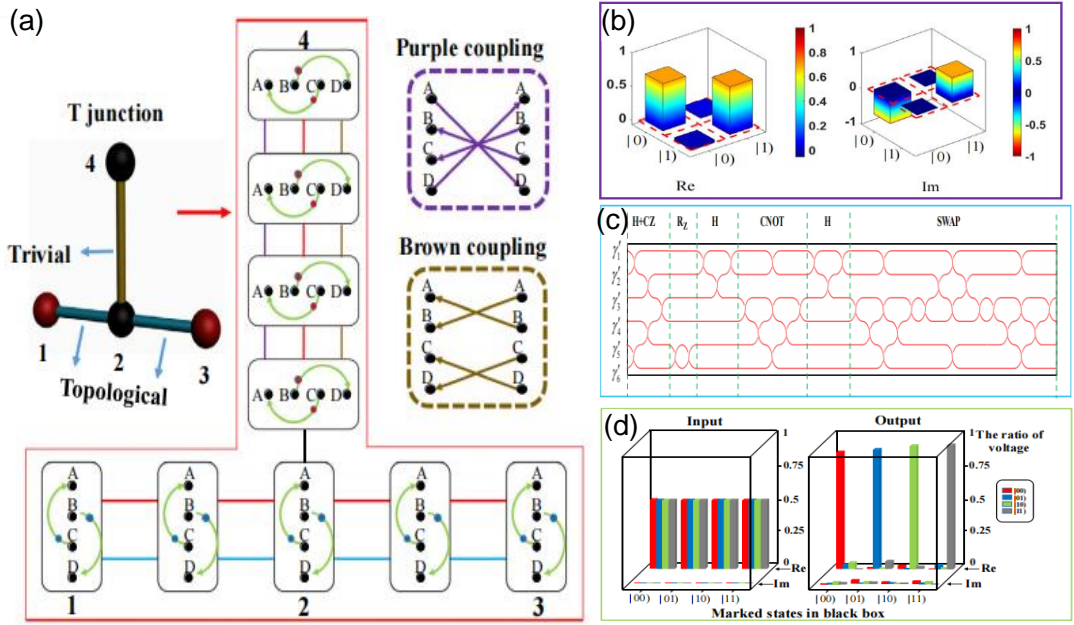

**Figure 11.** The circuit design and experimental results for the braiding process and Grover's algorithm. (a) Illustrations of a classical circuit for a T junction. (b) The output voltages of the braiding matrix. (c) The theoretical scheme of Grover's algorithm. (d) The inputs and outputs of Grover's algorithm. These figures are taken from [22]. Figures are adapted from CC-BY open access publications Wiley Online Library.

The designed circuit network for T junction is shown in Figure 11a. It composes three legs made of Kitaev chains (1-2, 2-3, and 2-4). The legs (1-2 and 2-3) can achieve the movement of the Majorana edge states by controlling resistance which determines the topological or trivial phase of the circuit segment. The leg (2-4) is a Kitaev chain with a variable phase $\phi$ which can achieve both the movement and phase rotation of topological edge states. The corresponding circuit Laplacian $J_{TR}$ for the T junction can be expressed as:

$$J_{TR} = \begin{pmatrix} \operatorname{Im} J_K(k) & -\operatorname{Re} J_K(k) \\ \operatorname{Re} J_K(k) & \operatorname{Im} J_K(k) \end{pmatrix}, \qquad (39)$$

where $J_K(k) = \dfrac{1}{2C}\begin{pmatrix} \varphi_k & 4iR_a^{-1}e^{-i\phi}\sin k \\ -4iR_a^{-1}e^{i\phi}\sin k & -\varphi_k \end{pmatrix}$ with $\varphi_k = -4R_b^{-1}\cos k - 2R_c^{-1}$. Here $\phi$ is circuit phase which determines the change of variable resistor. This circuit Laplacian corresponds to the Hamiltonian of the T junction. The whole braiding process can be described by a braiding matrix $U_{\gamma'_i \gamma'_j} = \begin{pmatrix} e^{-i\pi/4} & 0 \\ 0 & e^{i\pi/4} \end{pmatrix}$ in the qubit representation. It indicates the braiding for zero modes $\gamma'_i$ and $\gamma'_j$ which indicates a $\pi/2$ phase gate. Figure 11b shows the real and imaginary parts of output voltages in the experiment. $|0\rangle$ and $|1\rangle$ are defined to represent the initial states. It can be seen that the input states $|0\rangle$ and $|1\rangle$ are transformed into $1/\sqrt{2}(1-i)|0\rangle$ and $1/\sqrt{2}(1+i)|1\rangle$, respectively. It corresponds to the result of $\pi/2$ phase gate.

Based on the implementation of braiding operations in the above designed circuit, quantum algorithms can be further performed. Grover's search algorithm is one of the most important algorithms in quantum computation. Compared with classical algorithm, it is proved more efficient and can solve difficult problems. Now, the implementation of Grover's fast quantum search based on braiding operations in the circuit are introduced.

In Figure 11c, the braiding process for the two-qubit Grover's search algorithm is shown. The red lines exhibit the scheme of the braiding process. Six Majorana zero modes ($\gamma'_1$, $\gamma'_2$, $\gamma'_3$, $\gamma'_4$, $\gamma'_5$ and $\gamma'_6$) are used in the topological quantum computation scheme. Different braiding combinations can form different gate operations which is shown on the top of Figure 11c. Such a scheme for the topological quantum computation corresponds to the universal quantum route diagram one by one. Figure 11d shows the measured results of the output voltage ratio. The bottom shows four secretly marked states $|00\rangle$, $|01\rangle$, $|10\rangle$ and $|11\rangle$ in the black box. The ratio of the four states in the output voltage with only one evolution are shown by different colors of square cylinders. It can be seen that for each secretly marked state, only one evolution is needed to identify it, whereas classically three evaluations are needed in the worst case, and 2.25 evaluations are needed on average.

## 5. Quantum walk based on Classical Circuits

Quantum algorithm has been the focus of numerous studies and is expected to play an important role in future information processing, since it outperforms classical computation at many tasks [227-233]. Quantum walk is thought as one of the useful platforms to realize quantum algorithms. Based on quantum walk, many theoretical proposals have been put forward to realize quantum algorithms, which show the quantum speedups compared with the corresponding classical algorithms. Recently, based on photonic platforms, Qiang, et al. have shown that the experimental results about the quantum walk search algorithms [234], and Tang, et al have demonstrated the quantum fast hitting algorithm on an equivalent balanced tree [235]. However, this way cannot be easily extended to the general graphs and other quantum algorithms. Due to lack of controllability, the precisely control the distant couplings in the quantum walk is very hard, and the extensions to the high-scale experimental demonstrations are not possible. Therefore, how to display the scalability in the quantum walk based algorithms deserves to be explored. Recently, the appearance of electric circuit provides numerous advantages in showing the algorithms, due to the easy-control and scalability of the circuit network. Based on the circuit network, the quantum search on the graph and quantum fast hitting algorithms have been shown.

## 5.1. Electric-Circuit Realization of Fast Quantum Search

Quantum search algorithm is expected to play an important role in future information processing, since it outperforms classical algorithm. The quantum Grover's search algorithm can find the value of $w$ using of order $\sqrt{N}$ queries, which shows quadratic speedup to the fastest classical algorithm in searching unsorted databases. Such an algorithm is extremely important, both from fundamental and practical standpoints, because it is a basis of many other quantum algorithms.

Compared with the quantum Grover's algorithm, quantum search algorithms based on quantum walk are directly connected to the search in a physical database, which is related to the practical use closely. Lots of theoretic studies about the quantum search algorithms on the quantum walk have been provided. Based on the discrete-time quantum walk, Shenvi, et al. have performed a quantum search algorithm performs on a database of $N$ items with $O(\sqrt{N})$ calls to the oracle, yielding a speedup similar to other quantum search algorithms [236]. Childs, et al. have proposed an alternative search algorithm based on a continuous-time quantum walk on a graph [237, 238]. The quantum search algorithms have shown that on d-dimensional cube lattices, the search of the target vertex displays the quantum speedup with $O(\sqrt{N})$ at *d*>4, and $O(\sqrt{N}\log N)$ at *d*=4, but no quantum speedup at *d*<4. Then, by controlling the quantum walk on the two-dimensional lattice using an ancilla qubit, Tulsi and other researchers have shown that the quantum walk search algorithm can solve the problem in $O(\sqrt{N\log N})$ steps [239, 240]. Moreover, finding a marked edge or a marked complete subgraph in a complete graph based on quantum walk has also been discussed by Hillery, et al. [241]. Due to the special design evolution in the quantum search, Tulsi has focused on a general framework of the quantum search algorithms, in which some of specified transformations are replaced by the general unitary transformation [242].

However, although there are many quantum search algorithms based on quantum walk, it is generally believed that the high connectivity is the necessary condition to realize the quantum search. The belief is broken with Janmark, Meyer and Wong by a new design method involving the degenerate perturbation theory [243-245]. Furthermore, to overcome the structure of the graph causing the walker to be trapped, the quantum walk search algorithm is redesigned by increasing the weights of the link between the vertices [246, 247]. But, Philipp, et al. has shown that this quantum walk search algorithm meets the difficulty on the tree structures [248]. Recent study by Wang, et al. have shown that such a difficulty can be overcome if the edge weights can be adjusted [249].

Although these quantum walk related algorithms have been provided for decades, the experimental realizations of these algorithms are rare. Recently, Qiang, et al. have shown that the experimental results about the quantum walk search algorithms [244]. Three different kinds of graphs are provided, including five-vertex complete graph, the constructed six-vertex graph and fifteen-vertex graph via two-boson quantum walk. However, this way cannot be easily extended to the general graphs. Due to lack of controllability, the precisely control the distant couplings in the quantum walk is very hard, and the experimental demonstrations of quantum search for complex graphs are not possible.

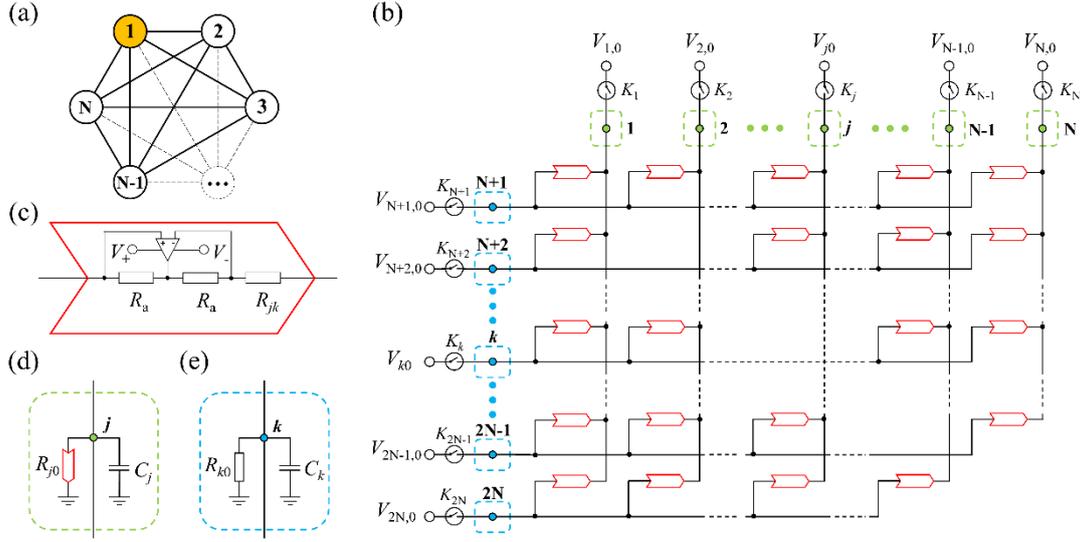

**Figure 12**. The experimental realizations of quantum walk search algorithms on the electric circuits. These figures are taken from [22]. Figures are adapted from CC-BY open access publications AAAS.

This problem of controllability can be overcome easily in the circuit, and the quantum search algorithms can be studied on the circuit networks. As shown in the Second Section, it is found that due to the good correspondence between the Kirchhoff's equation and the Schrodinger's equation, the electric circuit network can exhibit the evolution governed by the quantum dynamics. The changes of currents and voltages in the circuit network can correspond to the time evolution for the wave function of the microscopic world. Pan, et al. have shown that the complete graph with $N$ vertices in Figure 12, and realized the corresponding circuit for the CTQW search [22]. The correspondence between the circuit Laplacian and lattice Hamiltonian is provided below. In the quantum structure, the state functions of $N$ nodes contain real part and imaginary part. In order to simulate the evolution of these state functions in the classical circuit, $2N$ nodes are needed. The first $N$ nodes describe the real part of the state functions, and the last $N$ nodes correspond to the imaginary part. We use Kirchhoff's current law to derive the circuit equation

$$i\frac{dV_m}{dt} = \frac{i}{C_j}(-\frac{1}{R_{m0}} - \sum_{\langle n \rangle}\frac{1}{R_{mn}})V_j + \frac{i}{C_j}\sum_{\langle n \rangle}(\frac{1}{R_{mn}}V_n), \qquad (2)$$

where $V_m$ and $V_n$ are the voltages of the node $m$ and $n$. $C_{m0}$ and $R_{m0}$ represent the capacitance and the resistance between node $m$ and ground. $R_{mn}$ represent the resistance between node $m$ and $n$. $\langle n \rangle$ indicates the summation confined to other connected nodes. It can be reformulated in matrix form

$$i\partial_t |\phi(t)\rangle = G|\phi(t)\rangle, \tag{5}$$

where $|\phi(t)\rangle = (V_1(t), \cdots, V_{2N}(t))^T$. The matrix $G$ is

$$G = \frac{i}{C_0} \begin{pmatrix} -\frac{1}{R_{10}} - \sum_{\langle n \rangle} \frac{1}{R_{1,n}} & \cdots & 0 & \frac{1}{R_{1,N+1}} & \cdots & \frac{1}{R_{1,2N}} \\ \cdots & \cdots & \cdots & \cdots & \cdots & \cdots \\ 0 & \cdots & -\frac{1}{R_{N0}} - \sum_{\langle n \rangle} \frac{1}{R_{N,n}} & \frac{1}{R_{N,N+1}} & \cdots & \frac{1}{R_{2,2N}} \\ \frac{1}{R_{N+1,1}} & \cdots & \frac{1}{R_{N+1,N}} & -\frac{1}{R_{(N+1)0}} - \sum_{\langle n \rangle} \frac{1}{R_{N+1,n}} & \cdots & 0 \\ \cdots & \cdots & \cdots & \cdots & \cdots & \cdots \\ \frac{1}{R_{2N,1}} & \cdots & \frac{1}{R_{2N,N}} & 0 & \cdots & -\frac{1}{R_{2N0}} - \sum_{\langle n \rangle} \frac{1}{R_{2N,n}} \end{pmatrix}. \tag{6}$$

If we make $\frac{1}{C_0 R_{mn}} = -H_{m,n-N}$ and $\frac{1}{R_{m0}} = -\sum_{\langle n \rangle} \frac{1}{R_{mn}}$, the matrix $G$ can be express as

$$G = i \begin{pmatrix} 0 & -H_E \\ H_E & 0 \end{pmatrix}, \tag{7}$$

where

$$H_E = -\begin{pmatrix} \frac{1}{R_{1,N+1}} & \cdots & \frac{1}{R_{1,2N}} \\ \cdots & \cdots & \cdots \\ \frac{1}{R_{N,N+1}} & \cdots & \frac{1}{R_{2,2N}} \end{pmatrix}. \tag{8}$$

For such a RC circuit, the evolution of voltage corresponds to the Schrodinger equation $i\partial_t |\psi(t)\rangle = H|\psi(t)\rangle$ when the initial voltages are selected as $|\phi(0)\rangle = \begin{pmatrix} 1 \\ 0 \end{pmatrix} \otimes |\psi(0)\rangle$.

Because at a given time T, the voltages $|\phi(T)\rangle$ can be express as

$$|\phi(T)\rangle = e^{-iGT}|\phi(0)\rangle$$
$$= \begin{pmatrix} \cos(H_ET) & -\sin(H_ET) \\ \sin(H_ET) & \cos(H_ET) \end{pmatrix} \begin{pmatrix} |\psi(0)\rangle \\ O \end{pmatrix}. \quad (9)$$
$$= \begin{pmatrix} \cos(H_ET)|\psi(0)\rangle \\ \sin(H_ET)|\psi(0)\rangle \end{pmatrix}$$

When multiplication $(1 \; -i) \otimes I_N$ to the left, Eq. (9) can be written as

$$|\psi(T)\rangle = ((1 \; -i) \otimes I_N)|\phi(T)\rangle$$
$$= (\cos(H_ET) - i\sin(H_ET))|\psi(0)\rangle. \quad (10)$$
$$= e^{-iH_ET}|\psi(0)\rangle$$

So, the evolution of voltage in Eq. (5) corresponds to the Schrodinger equation $i\partial_t|\psi(t)\rangle = H|\psi(t)\rangle$, when $H_E$ corresponds to the Hamiltonian $H$.

Based on the circuit design above, the simulation results show that the time is coincidence with those obtained from quantum search algorithms, in which the time satisfies $t = \pi\sqrt{N}/2$ with $N = 4$. Furthermore, not only the one vertex can be found effectively in the circuit network, the search for more than one vertex can also be operated well in the electric circuit network. Ji, et al. have reported when the number of target vertices is $k$, and the optimal time find these $k$ vertices is also proportional to $\sqrt{N}$ where $N$ is the total number of vertices in the graph [250, 251].

## 5.2. Electric-Circuit Simulation of Quantum Fast Hitting with Exponential Speedup

Another important algorithm is the quantum fast hitting algorithm. The fast hitting algorithm concentrates the consuming time for a walker starting from an entrance vertex to the exit vertex. It has been proved that the classical algorithm needs the time in the exponential order of the height of tree to find the exit vertex. As a comparison, the quantum fast hitting algorithm based on quantum walk only requires the time in the polynomial order of the height to reach the exit. Therefore, the quantum fast hitting algorithm displays the exponential speedup over the classical algorithms. Childs, et al. have chosen the unbalanced tree to display the quantum fast hitting algorithm [252]. The tree $G_n$ is composed of two complete binary trees, and each one has the height $n+1$. The leaves of these two complete binary trees are connected with each other randomly. The total number of nodes for the tree is $N = 2^{n+2} - 2$, in which the entrance node is labeled by 1 and the exit node is $N$. When compared to the classical algorithms, the quantum fast hitting algorithm has introduced the quantum walk to finish the evolution process. As demonstrated by Childs, et al. [252], the quantum fast hitting algorithm displays the exponential quantum speedup when compare with the classical algorithm. However, the experimental implementation of quantum fast hitting algorithm is challenging, because the structure contains highly complex arrangements of an exponentially increasing number of nodes. Recently, the experimental demonstration of quantum fast hitting algorithm has been realized on the optical platform firstly [235, 253]. Not only in the waveguide system, the quantum fast hitting is also realized based on a full-stack photonic computing system [254]. Although the quantum fast hitting algorithm has been shown on the photonic platforms, the experimental implementation of the algorithm is verified in an equivalent way for the unbalanced tree. The direct experimental demonstration for the unbalanced tree is rather difficult on the photonic platform, because it requires the accurate couplings between two nodes further away.

However, this requirement can be easily overcome in the electric circuit [255]. Zhang, et al. have designed the circuit network for the unbalanced tree, which has been shown in Figure 13a.

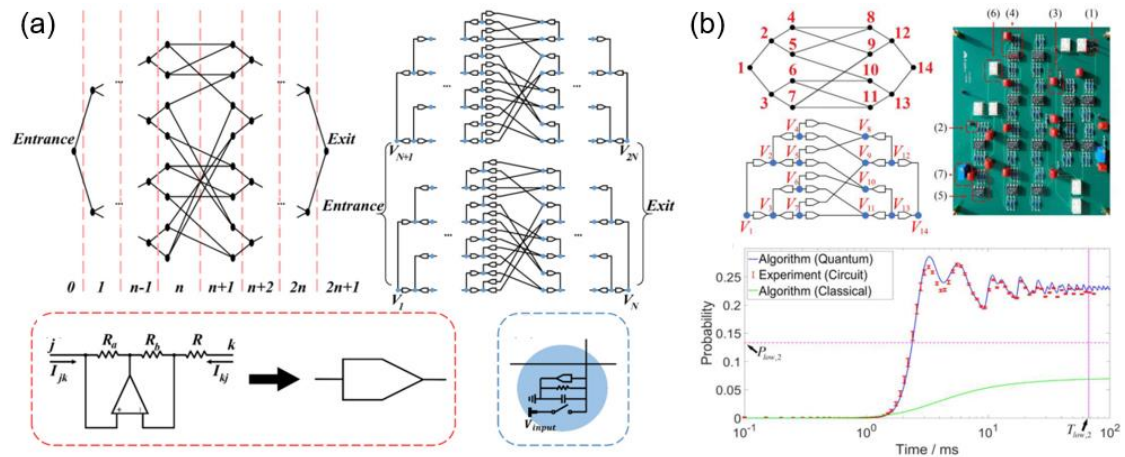

**Figure 13**. The realization of quantum fast hitting algorithms on the circuit network. (a) The construction design. At the bottom, the structures of electric components. (b) The graph of circuit for the unbalanced tree and the operation results of algorithm. These figures are taken

With the design, the numerical simulations and experimental results are shown in Figure 13b. From these results, it is also noted that when considering the probability to find the walker at the exit node, the quantum fast hitting algorithm gives the value which is exponentially larger that from the classical algorithm. Moreover, the circuit networks not only can display the speedup in the search and fast hitting problems, but also can show the advantage in the combinational logics.

## 6. Quantum combinational logics based on classical circuits

Combinatorial logic circuits consist of the modern digital circuit, and have been applied in many fields. Based on their operations, the functions of combinatorial logic circuits can be shown in three main areas. The first one is the arithmetic logic calculation, which contains many functions in mathematics. The second one is the data transmission, which includes the functions to control the data during the transmission. The third area is about code converter. Although the combinatorial logic circuit has these enormous applications, the processing speed of the combinatorial logic circuit is limited by the conventional structure in the classical circuit, which cannot provide a satisfied answer to the huge information processing nowadays.

As known in the combinatorial logic circuit, the negative-AND (NAND) gate has logical completeness, indicating that all logic functions in the circuit can be fulfilled with this NAND gate only. Obviously, the three main logic functions mentioned above can also be realized with the NAND gate. However, the requirement of fast processing speed still exists, and turns to the increase of operation speed for the NAND gate. Recently, Jensen, et al. have proposed a possible theoretic design based on the molecular structure [256]. The quantum NAND gate is mapped to the conjugated molecular system. However, the above proposal based on the molecular system is not realized easily, for it requires the dedicate control of molecular structure. Wang, et al. have demonstrated the quantum NAND-Tree algorithm on the photonic platform experimentally [257]. But it is hard to extend the construction to a larger system. The lack of adjustability in the photonic waveguide system greatly limits the application in a larger structure.

Recently, by utilizing the easy-control and scalability of the circuit network, the quantum NAND-Tree algorithm, quantum combinational logics, and related quantum two-player games have been shown.

**6.1. Quantum NAND-Tree algorithm on the circuit network**

The fast operation of the NAND gate has an important application in the NAND-Tree algorithm, which is commonly used in the detection of the working condition of the chip. The classical algorithm has shown that the optimal time is about $N^{0.753}$ with $N$ input bits to generate the output. As a comparison, Farhi, et al. have showed when the quantum coherence is designed into the algorithm, it only the time to generate the output in the order of $\sqrt{N}$, and the output result is 1 or 0 can exactly correspond to the output from the logic circuit [258]. Without doubt, the acceleration of the new NAND gate can lead to the acceleration of other types of logic gates in the circuit, and induce the quantum speedup of logic functions.

When referring to the section above, the good correspondence between the electric circuit network and the quantum Hamiltonian on the graph provides a new way to demonstrate the quantum speedup in the NAND-Tree algorithm and the combinatorial logic circuit [259]. Zhang, et al. have provided the circuit design for the quantum NAND-Tree, which has been provided in Figure 14a. Different inputs in the circuit can be controlled by the dual in-line package switch. The electric components connecting the electric nodes and the ground have been shown. The input 0 or 1 into the tree depends on the disconnection or connection between the top row and the leaf nodes of the tree. To realize the quantum algorithm, the wave function starts from the left part of chain, and then evolve into the right part and tree structure with time. The output is 0 if the probability of walker located outside the left part of chain is smaller than those inside the left part of chain. Otherwise, the output is 1.

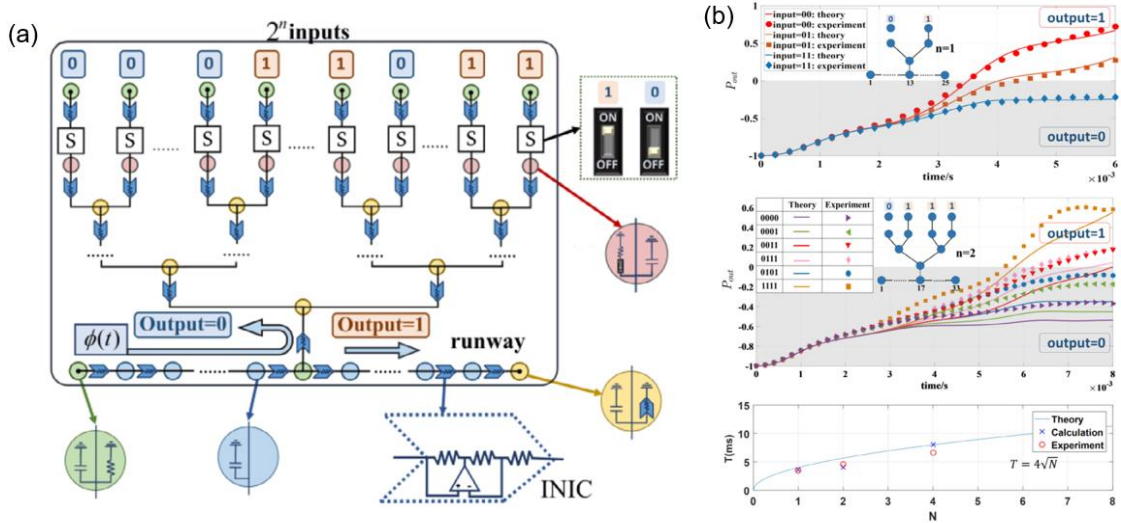

**Figure 14**. (a) The circuit design to realize the quantum NAND-Tree algorithm. (b) The experimental and simulation results for different layers of NAND-Trees. These figures are taken from [259]. Figures are adapted from CC-BY open access publications Wiley Online Library.

As shown in Figure 14b, the simulation and experiment results are provided for the quantum NAND-Tree with the depth $n=1$ and $n=2$. All these output results for different inputs coincide with the logic function of NAND-Tree algorithm. At the bottom of Figure 14b, the quantum speedup in the circuit network for the NAND-Tree algorithm has also been demonstrated. Therefore, the designed circuit networks realize the logic function of the NAND gates, and demonstrate the embodied quantum speedup of the quantum NAND-Tree algorithm.

### 6.2. Quantum combinational logics and their realizations with circuits

The review above has shown that the NAND gate can be designed to increase its operation speed, and display the quantum speedup in the quantum NAND-Tree algorithm. Considering the logical completeness of the NAND gate, other logic functions in the circuit can also be fulfilled with the NAND gate, and the operation speed can display quantum speedup. When considering the new design of the NAND gate has been realized on the circuit network, Tong, et al. have implemented the new adder, and the comparator and other logic operations in the same way [260]. At the top of Figure 15a, it provides the design for the half adder with the output as the carry bit. The graph structure includes two NAND gates, and each of them is labeled by the blue dashed frame. The function of the NAND gate is shown next to the half adder. In addition, the half adder with the sum bit is shown in the middle of Figure 15a. Besides the half adder, the graph structure of the comparator is shown at the bottom of Figure 15a.

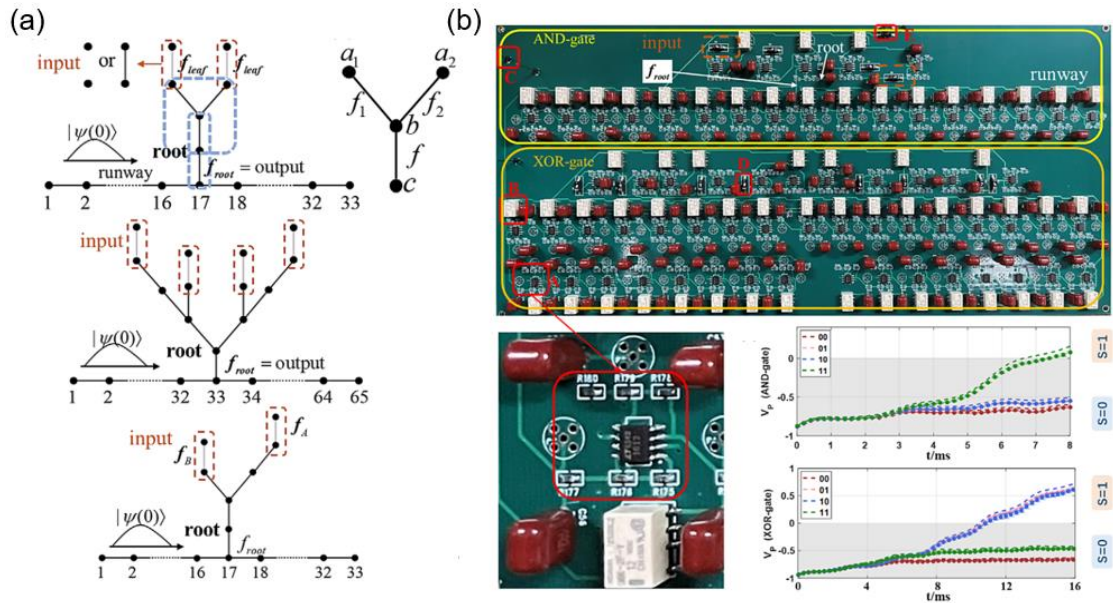

**Figure 15**. The designs for the logic gates. (a) The circuit design for the half adder and comparator. (b) The experimental results of the half adder. These figures are taken from [260]. Reproduced with permission [260] Copyright 2024, Wiley Online Library.

The experimental realization of half adder with the carry bit and sum bit has been addressed in Figure 15b. As shown in the bottom right of Figure 15b, the experimental outputs agree with the logic function of the half adder with the carry bit very well. Also, the experimental results for the half adder with the sum bit are shown. In this way, the functions of the half adder based on quantum walk have been implemented in the circuit network. Not only for the half adder, the comparator based on quantum walk has been realized in the circuit network.

The NAND gate not only works in the gate-based algorithm, but also operator as the basic element to realize some practical problems, for example, the game problem, etc.

### 6.3. Quantum Two-Player Games and Realizations with Circuits

Game theory studies the decision-making in different situations involving conflict and cooperation, with the aim of abstracting key elements of various competitive scenarios and scientifically investigating their characteristics [261-264]. The integrated artificial intelligence has been introduced into game theory problems, combining machine learning methods to excel in domains such as chess and Go, which results in the development of high-performance computer programs playing at a super-human level [265-272]. A fundamental example in the game problem is the two-player game, relating to the decisions for two competing agents. It is common to treat this problem into a decision tree, and judge the winner from the output of the tree. Therefore, the fast decision requires the quick calculation of the tree output. As shown in [273], if the two-player game tree is seen as the AND-OR tree, the minimax value of this tree has one-to-one correspondence to the solution of the AND-OR tree. In this way, the evaluation of game strongly depends on the optimal solution to the AND-OR tree. In the classical calculation of the AND-OR tree output [274, 275], the value of a balanced binary AND-OR tree without error is expected to obtain at the time in the order of $O(N^{\log_2[(1+\sqrt{33})/4]}) = O(N^{0.754})$.

Quantum algorithms provide the significant advantage in solving the specific two-player game. The simplest Prisoners' Dilemma problem is the one-step decision problem, which is proven without the dilemma with quantum strategies [276-278]. The many-step decision problems are also studied with quantum algorithms. By employing the idea of the Grover's search, a quadratic speedup can be implemented over naive classical algorithms [279-281]. Then, based on quantum circuit model, the bounded-error quantum algorithms were developed to calculating the output of the game tree [282, 283]. The query in the order of $O(\sqrt{N}\log N)$ is required to evaluate AND-OR formulas with size $N$.

Recently, Zhang, et al. have proposed a novel quantum scheme for a two-player game based on AND-OR tree structures [284]. They employ a subgame design technique to develop a quantum algorithm for the Hamiltonian AND-OR tree using continuous-time quantum walk. Their proposed algorithms achieve a query time complexity of $O(\sqrt{N})$ for evaluating preprocessed approximately balanced AND-OR trees. It not only exhibits quantum speedup, but also has speedup effects even compared to existing quantum algorithms. Furthermore, they validate the quantum speedup characteristics of this algorithm within circuit networks.

## 7. Quantum-inspired signal processing for implementing unitary

**transforms**

Unitary transform plays an important role in the area of signal processing which involves a wide range of applications [285-289], such as Hadamard transforms [290], z-transform [291, 292], operations involving wavelet transforms [293, 294], discrete Fourier transform to spectral analysis and filtering [295, 296]. Therefore, how to improve the efficiency of performing unitary transforms is important issue for current information processing, which have been involved in Refs. [297-299].

It is well known that quantum computation can offer advantageous information processing compared to classical counterparts, including unitary transformations. Hence, a direct idea is to extend the ideas to further enhance the strategies for performing unitary transforms. There are generally two well-known models for quantum computation, namely the circuit-based model and measurement-based model. For the former, computation is run by a sequence of unitary gates and displayed by its circuit diagram, which can narrate many popular algorithms [300-302]. In recent years, many works have been done to emulate the quantum circuit model based on classical circuits. The strategy in the above section 4 provides an important example of along the research line. Other examples include the schemes built on a programmable logic device [303, 304], the Hilbert-space-analog computing by analog microelectronic circuits [305], the quantum gates simulation by Field Programmable Gate Array (FPGA)-based circuit models [306], and the implementation of Deutsch-Jozsa algorithm by [307], etc [308-310].

For the latter, i.e. measurement-based quantum computing (MBQC) model [311], the computation is run by a sequence of single-qubit projective measurements on a prepared quantum graph state, which enables to simulate any quantum circuit in principle [312, 313]. It has been demonstrated that these two models are equivalent for universal quantum computation [312]. Even so, the corresponding work for the paradigm of MBQC in classical experimental systems has not been investigated much. An important work for starting the research along the line is proposed by Zhang, et al [314]. In their work, a microwave circuit is employed for MBQC, demonstrating the cases that corresponds to 8-qubit and 16-qubit graph state.

The main structure of the microwave circuit for simulating the graph state is quite simple. It is mainly composed of a microwave signal generator, a microwave signal receiver, and a processor whose main function is mixing the frequency of the revived microwave proposed by Ref. [314]. As shown in the upper panel of Figure 16 for four qubit classical microwave graph states (CMGS), the microwave signals with vertical and horizontal polarizations at a certain frequency are produced by a dual-polarized transmitting antenna, and received by four dual-polarized antennas. After an 8-channel microwave receiver, the signals are down-converted from a radio frequency to an intermediate frequency, and subsequently mixed by using an analog-to-digital converter (ADC) and digital signal processing (DSP) techniques. Finally, the signal $A$ obtained by applying specific band-pass filter to the measured one has the form, $A \propto \left[ \left( \hat{e}_{m1} | \left( \hat{e}_{m2} | \left( \hat{e}_{m3} | \left( \hat{e}_{m4} | \right] \right] | G \right)$, where $\hat{e}_{mj}$ denotes the projection direction and $j = 1,...,4$. The notation $(\ |\ )$ comes from the formulation in Ref. [314] and other relevant works, for showing that the measurements on the above microwave state has the similar math structure with the measurements on a quantum state. Then, the microwave state can be expressed by

$$|G\rangle \propto |h\rangle_1 |h\rangle_2 |h\rangle_3 |h\rangle_4 + |h\rangle_1 |h\rangle_2 |v\rangle_3 |v\rangle_4 + |v\rangle_1 |v\rangle_2 |h\rangle_3 |h\rangle_4 - |v\rangle_1 |v\rangle_2 |v\rangle_3 |v\rangle_4 \quad \text{(the}$$

normalization is omitted), which is the classical microwave analog of the four-photon polarization cluster state formally described by Ref. [315].

More properties of CMGS can be verified by properly setting the "local projections", or the direction of $\hat{e}_{mj}$. For example, like the quantum state tomography, one can also perform the projection on the CMGS according to the choice of basis and then plot the diagraph of the components corresponding to the density matrix. The results are given by the lower panel of the Figure 16. It is obviously that such a plot is exactly the same with those obtained by quantum states. The fidelity of the CMGS reported by the work is very high. From the setup, it can be noted that the it is relatively easy to scale up the system. In fact, in Ref. [314], the author has extended the device to simulate the 16-qubit quantum graph state.

The unitary transformation can be done by measuring the above CMGS state properly, inspired by the mathematical tool for proving equivalence of measurement-based quantum computing and quantum circuit model. As reported by Ref. [314], 2-by-2, 4-by-4, and 214-by-214 unitary transforms can be implemented by properly measuring the CMGS, which can be considered as the simulation of the corresponding measurement on single qubit, two-qubit, and 7-qubit quantum graph state respectively.

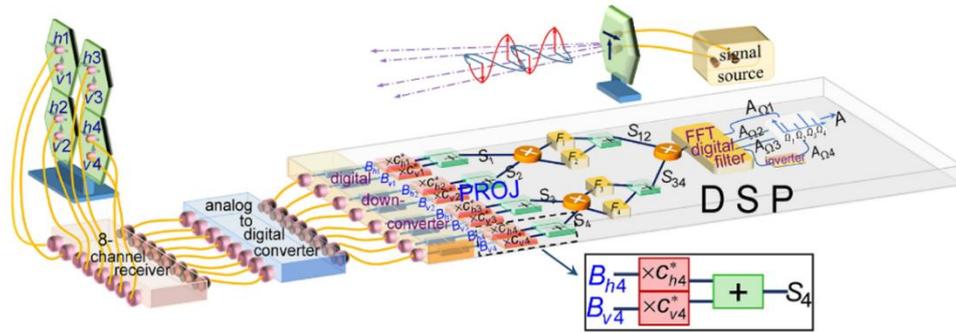

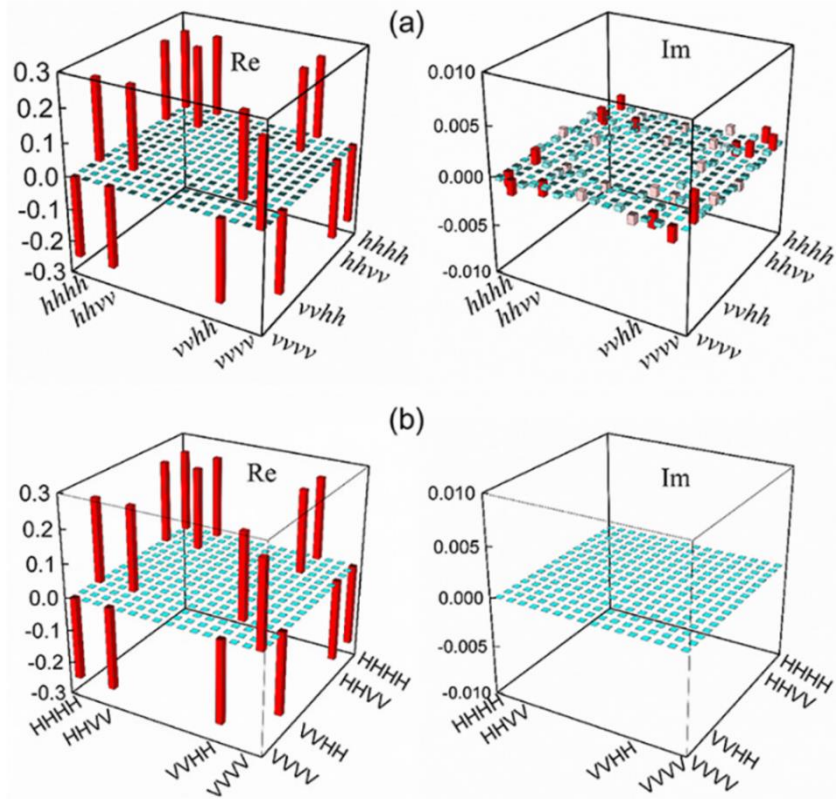

**Figure 16.** A designed classical microwave signal processing system to construct and characterize CMGS (upper panel). The density matrix of the 4-cebit classical cluster (lower panel). real part of the density matrix is in the left, and the imaginary part is in the right [314]. Figures are adapted from CC-BY open access publications Optical Society of America.

Moreover, the advantageous quantum algorithm can also be implemented, a typical example is the Grover algorithm. In Ref. [314], an example that corresponds to the two-qubit Grover search algorithm is given. From the experimental setup, it can be noticed that the whole implementation keeps in a simple and easy way, where only the measurement settings are required to be tuned, ensuring high stable and reliable outcomes. At present, microwave technology is widely used in many fields, such as radar detection [316, 317], electromagnetic countermeasure [318], electronic interference and communication [319, 320]. Accordingly, how to apply the design idea of CMGSs to improve the processing ability involving unitary transformations may be an alternative direction for future research.

## 8. Conclusions and Outlooks

After a decade of development, extensive research has been conducted in circuit systems to explore topological states of matter, investigate topological quantum computing, study quantum walk and quantum combinational logics, as well as develop quantum-inspired signal processing techniques for implementing unitary transforms. It has shown that electrical circuits manifest as a powerful platform in simulating quantum physics. Compared to other classical systems such as condensed matter materials, acoustics, optics, mechanics, cold atoms, and superconducting circuits, the advantages of classical circuits are expressed in the following three aspects.

The first aspect is to demonstrate advantages in implementing lattice systems. Classical circuits can implement a wide range of lattice couplings that are difficult or impossible to be achieved in other systems. These include arbitrary strength and distributions of disorder, non-Euclidean geometries, non-Abelian effects, high-dimensional spaces, and long-range non-local couplings. In contrast, systems like cold atoms or condensed matter materials are limited in their ability to simulate such diverse interactions due to the inherent constraints of their physical setups. The second thing is the flexibility in the design. Classical circuits offer a variety of electronic components — such as gain/loss elements, non-reciprocal couplings, nonlinear devices, and time-varying couplings—that can be combined in a highly flexible manner. These components enable the simulation of complex physical phenomena that are difficult to access in other systems. The third one is the easily-measured in the circuit. Classical circuits benefit from well-established measurement and fabrication technologies, which enable easy characterization in both the frequency and time domains. These systems operate in ambient conditions (room temperature and pressure), with relatively low noise interference, making them more accessible and practical for a wide range of experimental setups. In contrast, systems like cold atoms or superconducting circuits often require cryogenic temperatures or highly specialized environments, limiting their practical use for certain applications.

To sum up, since the classical circuit technology is relatively mature, circuit networks possess remarkable advantages of being non local coupling and reproducible signal, they can be used to study novel states of matter that are difficult or impossible to achieve in condensed matter systems and other systems. The more complex the problem, the more advantages can be demonstrated by utilizing circuits. Thus, more complex physical phenomena are expected to be explored using circuit networks in the future.

While, there are also some challenges in the circuits network. They do not naturally support quantum many-body correlations and interaction effects, and are currently limited to experimental realizations of few-body physics. Further exploration is needed on how to design circuits to simulate many-body strong correlation problems. Additionally, large-scale circuit systems need to be designed and implemented in order to achieve a wide range of applications. When the circuit size increases, classical circuits also face limitations in terms of losses and the stability of active components. The integration of the electronics industry offers an effective solution to this issue, wherein Complementary Metal Oxide Semiconductor (CMOS) process technology is employed as a substitute for PCB designs. Three main advantages on the chip-implemented circuit are explicitly shown below. The first advantage is the easily configurable components in the realization. Based on such a property, a more stable and reliable response for the output impedance is achieved through elaborately choosing the circuit components. The second thing is the enormous reduction of the parasitic effect within the chip. Generally speaking, in the conventional PCB platform, the magnitudes of this parasitic capacitance and the fore-end capacitance are similar, which hinders the application, e.g., the detection of ultra-small value capacitance, and so on. When referring to the chip, the parasitic capacitance can be nearly negligible, which greatly improves the capability in detecting weak physical quantities. The third advantage is the increase operating frequency of topological mode on the chip platform. Since the circuit components of the chip can be formulated exactly, the noise generated on the impedance spectrum is easily forecasted. Considering three main advantages above, the new designs reported in the PCB platforms can be realized on the chips without doubt. Moreover, although we only discuss the physical principle based on PCB here, it is also possible to demonstrate the quantum speedup on the chip for more complex problems, including games, economics, cybersecurity, computer science, and finance, among others. Many more new circuit devices with powerful functions, especially for practical fast information processing, are expected to be fabricated to serve society.

## Acknowledgements

This work was supported by the National key R & D Program of China under Grant No. 2022YFA1404900 and the National Natural Science Foundation of China (12234004).

## Conflict of Interest

The authors declare no conflict of interest.

## Keywords